# State Feedback Controllers with Operational Constraints

Eugene Lavretsky

*Abstract* – In this paper, a state feedback control design with min/max operational limiting constraints is developed for multi-input-multi-output linear time invariant systems. Specifically, servo-tracking control problems with input and output constraints are considered. For static servo-controllers, the output design limits are imposed component-wise on the system selected output, which is of the same dimension as the control input. For dynamic servo-controllers, operational constraints are applied to the system inputs and outputs. The proposed control solution also includes an anti-windup protection logic for dynamic servo-controllers with integral action. The developed method is based on the Nagumo Theorem for forward invariance, the Comparison Lemma for inclusion of input/output inequality constraints, and on the min-norm optimal controllers for synthesis. The derived design is similar and directly related to the method of Control Barrier Functions. Simulation trade studies are presented to illustrate benefits of the proposed control methodology for aerial flight critical systems.

## 1. Introduction and Problem Formulation

Consider the controllable Multi-Input-Multi-Output (MIMO) Linear Time Invariant (LTI) dynamical system,

$$\text{Open-Loop System Dynamics}: \dot{x} = Ax + Bu, \quad x \in R^n$$
$$\text{Limited Output}: y_{\lim} = C_{\lim} x, \quad y_{\lim} \in R^m \quad (1.1)$$

where $x \in R^n$ is the $n-$dimensional state vector, $u \in R^m$ is the $m-$dimensional control input, and $y_{\lim} \in R^m$ is the system $m$–dimensional vector of limited outputs to be kept within the desired min/max bounds $\left(y_{\lim}^{\min}, y_{\lim}^{\max}\right) \in R^m$, component-wise.

In (1.1), the system matrices $(A, B, C_{\lim})$ are of the corresponding dimensions and the matrix pair $(A, B)$ is controllable. It is further assumed that the system state vector $x$ is accessible for control design, as the system output measurement.

Of interest is the control design with min/max ("box") component-wise output constraints. Specifically, a state feedback control input $u$ needs to be found such that the closed-loop system is stable and the limited output $y_{\lim}$ evolves within the predefined min/max operational constraint bounds, component-wise.

$$y_{\lim}^{\min} \leq y_{\lim} \leq y_{\lim}^{\max} \quad (1.2)$$

If such a controller can be designed then operational constraints (1.2) become "soft" to distinguish them from the "hard" constraints that are typically represented by the static saturation function,

$$\text{sat}_{y_{\min}}^{y_{\max}}(y_{\lim}) = \max\left(y_{\min}, \min(y_{\lim}, y_{\max})\right) \quad (1.3)$$

where min/max operations are applied component-wise on the system output $y_{\lim}$.

This paper presents derivations of state feedback controllers with soft operational constraints (1.2) that are achieved and enforced by feedback connections in order to preserver stability, boundedness and robustness of the corresponding closed-loop system trajectories. The developed control methodology applies to both stable and unstable open-loop MIMO LTI systems, with possible extensions to the class of nonlinear affine-in-control dynamics.

Let $u_{bl}$ denote a baseline controller for (1.1), designed without an explicit consideration of the output limits. For example, such a controller could represent a state feedback for stabilization or a servo-controller with command-proportional feedforward terms, for tracking external bounded commands. Dynamic servo-controllers are also possible and their design will be considered later in the paper.

In order to explicitly account for the operational constraints (1.2), the system control input is defined as,

$$u = u_{bl} + \pi \tag{1.4}$$

where $u_{bl} \in R^m$ is the baseline controller and $\pi \in R^m$ is an augmentation policy that will be designed to enforce soft limits on the system output. Control definition (1.4) yields the closed-loop system dynamics,

$$\text{Closed-Loop System Dynamics}: \dot{x} = Ax + B\underbrace{(u_{bl} + \pi)}_{u} \tag{1.5}$$

$$\text{Limited Output}: y_{\lim} = C_{\lim} x$$

with the state feedback control augmentation signal $\pi$ to be designed such that the operational constraints (1.2) are satisfied and enforced via feedback connections.

## 2. Constrained Quadratic Program for Control Augmentation Design

The output operational constraints (1.2) can be viewed as a subset (not necessarily bounded) within the system state space and the problem of enforcing these limits reduces to attaining forward invariance of the corresponding closed-loop system trajectories with respect to the predefined set. The Nagumo Theorem [6] formulates a criterion to achieving forward invariance for a dynamical system. It requires the system velocity vector to point inside of the limiting set at the set boundary. The Nagumo theorem forms the basis for the control augmentation design reported in this paper.

Motivated by the min-norm controller design method [4], consider the following Quadratic Program (QP) [5], with linear constraints (1.2).

$$\text{Minimization Cost}: J(\pi) = (\pi^T R_\pi \pi) \to \min_\pi$$

$$\text{Output Constraints}: y_{\lim}^{\min} \leq \underbrace{(C_{\lim} x)}_{y_{\lim}} \leq y_{\lim}^{\max} \tag{2.1}$$

or equivalently

$$\text{Minimization Cost}: J(\pi) = (\pi^T R_\pi \pi) \to \min_\pi$$

$$\text{Output Constraints}: \begin{cases} h_1(x) = y_{\lim}^{\min} - C_{\lim} x \leq 0 \\ h_2(x) = C_{\lim} x - y_{\lim}^{\max} \leq 0 \end{cases} \tag{2.2}$$

Prior to solving QP (2.2), the output constraints need to be modified to become directly dependent on the control augmentation policy $\pi$. That is a standard modification in constrained optimization problems [5]. The method for constraints modification depends on the vector relative degree of the system output with respect to its control input.

### A. Vector Relative Degree for MIMO LTI Systems

**Definition 1.** The system $m$-dimensional limited output $y_{\lim}$ (1.5) has vector relative degree $r = (r_1 \ \ldots \ r_m)$ with respect to the system control augmentation input $\pi$, where $1 \leq r_i \leq n, \forall 1 \leq i \leq m$, if $(r_i)_{i=1,\ldots,m}$ is the least number of times one has to differentiate the $i^{th}$ output component $(y_{\lim})_i$ to have at least one of the $m$ inputs $(u_j)_{j=1,\ldots,m}$ appear explicitly [10],

$$\left[ \left\| \nabla_u \left( (y_{\lim})_i^{(k)} \right) \right\| = 0, \forall 0 \leq k \leq r_i - 1 \right] \wedge \left[ \left\| \nabla_u \left( (y_{\lim})_i^{(r_i)} \right) \right\| \neq 0 \right], \quad i = 1,\ldots,m \tag{2.3}$$

and the control-to-output sensitivity matrix $H_u \in R^{m \times m}$ is nonsingular.

$$\det H_u = \det \begin{pmatrix} \nabla_u\left((y_{\lim})_1^{(r_1)}\right) \\ \nabla_u\left((y_{\lim})_2^{(r_2)}\right) \\ \vdots \\ \nabla_u\left((y_{\lim})_m^{(r_m)}\right) \end{pmatrix} \neq 0 \tag{2.4}$$

Based on (2.3), for MIMO LTI systems such as (1.5), the output relative degree relations (2.3), (2.4) can be written explicitly in terms of the system parameters.

$$\begin{pmatrix} (y_{\lim})_1^{(r_1)} \\ (y_{\lim})_2^{(r_2)} \\ \vdots \\ (y_{\lim})_m^{(r_m)} \end{pmatrix} = \begin{pmatrix} (C_{\lim})_1 A^{r_1} \\ (C_{\lim})_2 A^{r_2} \\ \vdots \\ (C_{\lim})_m A^{r_m} \end{pmatrix} x + \underbrace{\begin{pmatrix} (C_{\lim})_1 A^{r_1-1} B \\ (C_{\lim})_2 A^{r_2-1} B \\ \vdots \\ (C_{\lim})_m A^{r_m-1} B \end{pmatrix}}_{H_u \in \mathbb{R}^{m \times m}} u \tag{2.5}$$

In what follows, input-to-output representations similar to (2.5) will be utilized to derive analytic expressions for control augmentation state feedback policies to enforce the desired output operational constraints (1.2) component-wise.

*B. Output Constraints Modification*

The output constraints (1.2) are modified to embed the forward invariance criterion from the Nagumo Theorem.

For a fixed $i = 1, \ldots, m$, suppose that $r_i \geq 1$. Consider a stable polynomial of order $r_i$, with the real roots $\{\lambda_{ij}\}_{j=1,\ldots,r_i}$ located in the open left half complex plane, $\lambda_{ij} \in \mathbb{C}^-$,

$$\phi_i(s) = \prod_{j=1}^{r_i}(s - \lambda_{ij}) = \sum_{j=0}^{r_i} c_{ij} s^j \tag{2.6}$$

where $c_{ij}$ denotes the $j^{th}$ coefficient of the $i^{th}$ polynomial. Clearly, the first and the last coefficients of $\phi_i(s)$ can be computed explicitly,

$$c_{i0} = \prod_{j=1}^{r_i}(-\lambda_{ij}), \quad c_{ir_i} = 1 \tag{2.7}$$

and by the definition, $c_{i0} > 0$, for every $i = 1, \ldots, m$.

**Lemma 1.** For the system (1.1), suppose that $y_{\lim}$ has vector relative degree $r = (r_1 \ \ldots \ r_m)$, with $r_i \geq 1$ for all $i = 1, \ldots, m$. Consider the modified output and its respective modified output constraints,

$$Y_{\lim} = \underbrace{\begin{pmatrix} \phi_1(s) & \cdots & 0 \\ \vdots & \ddots & \vdots \\ 0 & \cdots & \phi_m(s) \end{pmatrix}}_{\Phi(s)} y_{\lim} = \Phi(s) y_{\lim}$$

$$H(x, u_{bl}, \pi) = \begin{pmatrix} \Phi(s)\underbrace{(y_{\lim}^{\min} - y_{\lim})}_{h_1(x)} \\ \Phi(s)\underbrace{(y_{\lim} - y_{\lim}^{\max})}_{h_2(x)} \end{pmatrix} = \begin{pmatrix} \Phi(s) h_1(x) \\ \Phi(s) h_2(x) \end{pmatrix} \leq 0 \tag{2.8}$$

where $(h_1(x), h_2(x))$ are the original output constraint functions from (2.2), with stable polynomials $\phi_i(s)$ (2.6), treated as differentiation operators with respect to $s$. Then,

$$Y_{\lim} = H_x x + H_u (u_{bl} + \pi)$$

$$\underbrace{H(x, u_{bl}, \pi)}_{\begin{pmatrix} H_1(x, u_{bl}, \pi) \\ H_2(x, u_{bl}, \pi) \end{pmatrix}} = \begin{pmatrix} -Y_{\lim} + \alpha_\pi y_{\lim}^{\min} \\ Y_{\lim} - \alpha_\pi y_{\lim}^{\max} \end{pmatrix} = \begin{pmatrix} -I_m \\ I_m \end{pmatrix} H_u \pi + \underbrace{\begin{pmatrix} -H_x x - H_u u_{bl} + \alpha_\pi y_{\lim}^{\min} \\ H_x x + H_u u_{bl} - \alpha_\pi y_{\lim}^{\max} \end{pmatrix}}_{\Delta H(x, u_{bl}) = \begin{pmatrix} \Delta H_1(x, u_{bl}) \\ \Delta H_2(x, u_{bl}) \end{pmatrix}} \leq 0 \qquad (2.9)$$

where $Y_{\lim} = \Phi(s) y_{\lim}$ is the system modified limited output, $\Phi(s) \in R^{m \times m}$ is a diagonal matrix of stable polynomials (2.6), with $\phi_i(s)$ on its $i^{th}$ diagonal,

$$H_x = \begin{pmatrix} (C_{\lim})_1 \phi_1(A) \\ \vdots \\ (C_{\lim})_m \phi_m(A) \end{pmatrix} = \begin{pmatrix} (C_{\lim})_1 \prod_{j=1}^{r_1} (A - \lambda_{1j} I_n) \\ \vdots \\ (C_{\lim})_m \prod_{j=1}^{r_m} (A - \lambda_{mj} I_n) \end{pmatrix}, \quad H_u = \begin{pmatrix} (C_{\lim})_1 A^{r_1-1} B \\ (C_{\lim})_2 A^{r_2-1} B \\ \vdots \\ (C_{\lim})_m A^{r_m-1} B \end{pmatrix} \qquad (2.10)$$

are the system state and control sensitivity matrices respectively, and $\alpha_\pi \in R^{m \times m}$ is a diagonal matrix, with its positive diagonal elements defined as the zero-order coefficients of the $i^{th}$ polynomial $\phi_i(s)$ (2.6).

$$\alpha_\pi = \begin{pmatrix} c_{10} & \cdots & 0 \\ \vdots & \ddots & \vdots \\ 0 & \cdots & c_{m0} \end{pmatrix}, \quad c_{i0} = \prod_{j=1}^{r_i} (-\lambda_{ij}) > 0, \quad \forall i = 1, \ldots, m \qquad (2.11)$$

***Proof of Lemma 1.***

Applying $\phi_i(s)$ to the $i^{th}$ components of the two constraint functions $h_1(x)$ and $h_2(x)$ from (2.2), gives

$$H_{1i}(x, u_{bl}, \pi) = \phi_i(s) h_{1i}(x_p) = \sum_{j=0}^{r_i} c_{ij} \left( y_{\lim}^{\min} - y_{\lim} \right)_i^{(j)} = c_{i0} \left( y_{\lim}^{\min} \right)_i - \underbrace{\sum_{j=0}^{r_i} c_{ij} \left( y_{\lim} \right)_i^{(j)}}_{(Y_{\lim})_i}$$

$$H_{2i}(x, u_{bl}, \pi) = \phi_i(s) h_{2i}(x_p) = \sum_{j=0}^{r_i} c_{ij} \left( y_{\lim} - y_{\lim}^{\max} \right)_i^{(j)} = -c_{i0} \left( y_{\lim}^{\max} \right)_i + \underbrace{\sum_{j=0}^{r_i} c_{ij} \left( y_{\lim} \right)_i^{(j)}}_{(Y_{\lim})_i} \qquad (2.12)$$

and the $i^{th}$ modified output component $(Y_{\lim})_i$ in (2.12) can be computed explicitly in terms of the system parameters.

$$(Y_{\lim})_i = \sum_{j=0}^{r_i} c_{ij} \left( y_{\lim} \right)_i^{(j)} = c_{ir_i} \left( y_{\lim} \right)_i^{(r_i)} + \sum_{j=1}^{r_i - 1} c_{ij} \left( y_{\lim} \right)_i^{(j)}$$

$$= \underbrace{\left[ (C_{\lim})_i A^{r_i} x + (C_{\lim})_i A^{r_i - 1} B u \right]}_{c_{ir_i} (z_{\lim})_i^{(r_i)}} + (C_{\lim})_i \left( \sum_{j=1}^{r_i - 1} c_{ij} A^j \right) x \qquad (2.13)$$

$$= \underbrace{\left[ (C_{\lim})_i A^{r_i - 1} B \right]}_{(H_u)_i} (u_{bl} + \pi) + (C_{\lim})_i \underbrace{\left( \sum_{j=0}^{r_i} c_{ij} A^j \right)}_{\phi_i(A)} x = (H_u)_i \pi + (H_x)_i x - (H_u)_i u_{bl}$$

$$\underbrace{\phantom{(C_{\lim})_i \left( \sum_{j=0}^{r_i} c_{ij} A^j \right)}}_{(H_x)_i}$$

In (2.13),

$$\phi_i(A) = \sum_{j=0}^{r_i} c_{ij} A^j = \prod_{j=1}^{r_i} (A - \lambda_{ij} I_n) \tag{2.14}$$

is the matrix polynomial with the same coefficients as in (2.6). Therefore, calculation of the polynomial coefficients $c_{ij}$ is not required to define the modified output signal.

$$Y_{\lim} = H_x x + H_u \underbrace{(u_{bl} + \pi)}_{u} \tag{2.15}$$

Substituting (2.13) into (2.12), results in

$$\begin{aligned} H_{1i}(x, u_{bl}, \pi) &= c_{i0}(y_{\lim}^{\min})_i - (Y_{\lim})_i = -(H_u)_i(u_{bl} + \pi) - (H_x)_i x + c_{i0}(y_{\lim}^{\min})_i \\ H_{2i}(x, u_{bl}, \pi) &= -c_{i0}(y_{\lim}^{\max})_i + (Y_{\lim})_i = (H_u)_i(u_{bl} + \pi) + (H_x)_i x - c_{i0}(y_{\lim}^{\max})_i \end{aligned} \tag{2.16}$$

where $c_{i0}$ is from (2.7). Define a strictly positive-definite diagonal matrix,

$$\alpha_\pi = \begin{pmatrix} c_{10} & \cdots & 0 \\ \vdots & \ddots & \vdots \\ 0 & \cdots & c_{m0} \end{pmatrix} > 0 \tag{2.17}$$

and rewrite (2.16) in a vector form.

$$\begin{aligned} H(x, u_{bl}, \pi) &= \begin{pmatrix} H_1(x, u_{bl}, \pi) \\ H_2(x, u_{bl}, \pi) \end{pmatrix} = \begin{pmatrix} -I_m \\ I_m \end{pmatrix} (H_u(u_{bl} + \pi) + H_x x) + \begin{pmatrix} \alpha_\pi y_{\lim}^{\min} \\ -\alpha_\pi y_{\lim}^{\max} \end{pmatrix} \\ &= \begin{pmatrix} -I_m \\ I_m \end{pmatrix} H_u \pi + \underbrace{\begin{pmatrix} -H_x x - H_u u_{bl} + \alpha_\pi y_{\lim}^{\min} \\ H_x x + H_u u_{bl} - \alpha_\pi y_{\lim}^{\max} \end{pmatrix}}_{\Delta H(x, u_{bl})} = \begin{pmatrix} -I_m \\ I_m \end{pmatrix} H_u \pi + \Delta H(x, u_{bl}) \end{aligned} \tag{2.18}$$

Therefore, the modified output constraints are defined as,

$$H(x, u_{bl}, \pi) = \begin{pmatrix} H_1(x, u_{bl}, \pi) \\ H_2(x, u_{bl}, \pi) \end{pmatrix} = \begin{pmatrix} -I_m \\ I_m \end{pmatrix} H_u \pi + \Delta H(x, u_{bl}) \tag{2.19}$$

with the corresponding $i^{th}$ rows written explicitly in terms of the system parameters, for all $i = 1, \ldots, m$.

$$\begin{aligned} (H_u)_i &= (C_{\lim})_i A^{r_i - 1} B \\ \Delta H_{1i}(x, u_{bl}) &= -(C_{\lim})_i \prod_{j=1}^{r_i} (A - \lambda_{ij} I_n) x - (H_u)_i u_{bl} + \left( \prod_{j=1}^{r_i} (-\lambda_{ij}) \right) (y_{\lim}^{\min})_i \\ \Delta H_{2i}(x, u_{bl}) &= (C_{\lim})_i \prod_{j=1}^{r_i} (A - \lambda_{ij} I_n) x + (H_u)_i u_{bl} - \left( \prod_{j=1}^{r_i} (-\lambda_{ij}) \right) (y_{\lim}^{\max})_i \end{aligned} \tag{2.20}$$

The proof of Lemma 1 is complete. □

If the $i^{th}$ output component $(z_{\lim})_i$ has relative degree $r_i$ then (2.8) requires that the following vector differential inequality of order $r_i$ takes place,

$$\phi_i(s) \begin{pmatrix} h_{1i}(x) \\ h_{2i}(x) \end{pmatrix} \leq 0 \tag{2.21}$$

where $\phi_i(s)$ is a stable polynomial differential operator of order $r_i$. The modified output constraints (2.9), (2.10) represent vector differential inequalities and as such, forward invariance of the related limiting subsets in the system state space needs to be analyzed.

## C. Forward Invariance Lemma

Lemma 2 gives sufficient conditions for forward invariance of a set defined by a scalar output whose relative degree is greater or equal than one. It is a direct corollary from the Nagumo Theorem [6], [7], [14].

**Lemma 2.** Consider an $n-$dimensional system of ordinary differential equations,

$$\dot{x} = f(x), \quad x \in R^n \tag{2.22}$$

with a Lipchitz in $x$ vector function $f : R^n \to R^n$. Let $h : R^n \to R$ be an $r-$times continuously differentiable scalar function of a vector argument. Consider the subset in $R^n$,

$$H = \{x \in R^n : h(x) \le 0\} \tag{2.23}$$

and assume that it is not empty. Suppose that for all initial conditions $x(0) = x_0 \in H$ the system trajectories are well-defined forward in time.

Let $\phi_r(s) = \prod_{j=1}^{r}(s - \lambda_j) = \sum_{j=0}^{r} c_j s^j$ denote a stable polynomial of order $r \ge 1$, with real eigenvalues $\{\lambda_j\}_{j=1,\ldots,r}$ in the open left-hand complex plane. Consider the set

$$\tilde{H}_r = \{x \in R^n : \tilde{h}_r(x) = \phi_r(s)h(x) \le 0\} \tag{2.24}$$

where

$$\tilde{h}_r(x) = \phi_r(s)h(x) = \sum_{j=0}^{r} c_j h^{(j)}(x(t)) \tag{2.25}$$

and $h^{(j)}(x(t))$ is the $j^{th}$ time derivative of $h(x)$ along the system trajectories. Then $\tilde{H}_r$ is forward invariant (FI) with respect to the system trajectories is and only if $H$ is FI.

$$\left[x(t) \in \tilde{H}_r\right] \Leftrightarrow \left[x(t) \in H\right], \forall t \ge 0 \tag{2.26}$$

***Proof of Lemma 2.*** *Sufficiency*: By the Nagumo Theorem, the set $H$ is FI with respect to the system dynamics (2.22) if an only if the state derivative vector $f(x)$ points inside the set at the set boundary.

$$\left[x(t) \in H, \forall t \ge 0\right] \Leftrightarrow \left[\dot{h}(x)\big|_{h(x)=0} = \left(\frac{\partial h(x)}{\partial x} f(x)\right)_{h(x)=0} \le 0\right] \tag{2.27}$$

For $r = 1$, if $\tilde{H}_1$ is FI then

$$\tilde{h}_1(x) = \phi_1(s)h(x) = (s - \lambda_1)h(x) = \dot{h}(x) - \lambda_1 h(x) \le 0 \tag{2.28}$$

Consequently,

$$\dot{h}(x)\big|_{h(x)=0} = \tilde{h}_1(x)\big|_{h(x)=0} = \left(\dot{h}(x) - \lambda_1 h(x)\right)_{h(x)=0} \le 0 \tag{2.29}$$

and so, by the Nagumo criterion (2.27), the set $H$ is also FI.

$$\tilde{H}_1 \subset H \tag{2.30}$$

For $r = 2$, if $\tilde{H}_2$ is FI then

$$\tilde{h}_2(x) = (s - \lambda_2)\underbrace{(s - \lambda_1)h(x)}_{\tilde{h}_1(x)} = (s - \lambda_2)\tilde{h}_1(x) \leq 0 \qquad (2.31)$$

and therefore, $\tilde{H}_1$ is FI, which in turn implies $H$ is FI.

$$\tilde{H}_2 \subset \tilde{H}_1 \subset H \qquad (2.32)$$

Continuation of this argument by induction, gives

$$\tilde{H}_r \subset \tilde{H}_{r-1} \subset \ldots \subset \tilde{H}_1 \subset H \qquad (2.33)$$

and proves the sufficiency condition of the lemma.

*Necessity*: This property is proved by the contradiction argument. For $r = 1$, suppose that $H$ is FI but $\tilde{H}_1$ is not. Since $H$ is FI then by Nagumo's Theorem, (2.27) takes place and consequently (2.28) is true on the set boundary $h(x) = 0$, and therefore $\tilde{H}_1$ is FI, which is a contradiction to the argument. Continuing the induction argument on the relative degree, yields necessity of (2.26) and completes the proof of Lemma 2. □

*D. Quadratic Programming Minimization Problem Formulation*

Lemma 2 provides sufficient condition for enforcing forward invariance when the system limited output components have relative degree greater than or equal to one. Based on that fact and using the modified input-output constraints (2.19), consider the following QP formulation.

$$\begin{aligned} \text{Minimizaion Cost}: J(\pi) = (\pi^T R_\pi \pi) &\to \min_w \\ \text{Output Constraints}: H(x, u_{bl}, \pi) = \begin{pmatrix} -I_m \\ I_m \end{pmatrix} H_u \pi + \Delta H(x, u_{bl}) &\leq 0 \end{aligned} \qquad (2.34)$$

Note that in (2.34) the minimization cost is quadratic and the constraints are linear, both with respect to the control decision policy $\pi$ and that is the enabler for deriving the QP analytical solution, which is typical of any optimization-based control formulations with constraints, including but not limited to linear quadratic optimal control and Pontryagin Principle of Maximum [3].

Since the minimization cost is quadratic and the constraint functions are linear, QP (2.34) has the unique optimal solution policy $\pi^*$, [5], which will be derived analytically in the next section.

## 3. QP-Based Control Augmentation Design with Min/Max Operational Constraints

Given the modified QP formulation in (2.34), consider the corresponding Lagrangian function,

$$L(x, u_{bl}, \pi, \lambda) = J(\pi) + \lambda^T H(x, u_{bl}, \pi) = \pi^T R_\pi \pi + \lambda^T \left( \begin{pmatrix} -I_m \\ I_m \end{pmatrix} H_u \pi + \Delta H(x, u_{bl}) \right) \qquad (3.1)$$

with the Lagrange multiplier vector-coefficients.

$$\lambda = \begin{pmatrix} \lambda_1 \\ \lambda_2 \end{pmatrix} \in R^{2m}, \quad \lambda_k \in R^m, \quad k = 1, 2 \qquad (3.2)$$

With respect to the control decision variable $\pi$, the Lagrangian (3.1) is convex, quadratic and differentiable. Therefore, Karush-Kuhn-Tucker (KKT) conditions for optimality are applicable for any $x \in R^n$ and $u_{bl}, \pi \in R^m$, [5].

$$\text{Stationarity}: \frac{\partial L(x, u_{bl}, \pi, \lambda)}{\partial \pi} = 0$$
$$\text{Primal Feasibility}: H(x, u_{bl}, \pi) \leq 0 \quad (3.3)$$
$$\text{Dual Feasibility}: \lambda \geq 0$$
$$\text{Complementary Slackness}: \lambda_i H_i(x, u_{bl}, \pi) = 0, \quad i = 1, 2, \ldots, (2m)$$

Solving the KKT stationarity conditions,

$$\frac{\partial L(x, u_{bl}, \pi, \lambda)}{\partial \pi} = 2 R_\pi \pi - H_u^T \begin{pmatrix} I_m & -I_m \end{pmatrix} \lambda = 0 \quad (3.4)$$

for the optimal decision policy $\pi^*$, gives

$$\pi^* = 0.5 R_\pi^{-1} H_u^T (\lambda_1 - \lambda_2) \quad (3.5)$$

To compute the optimal Lagrange coefficients $(\lambda_1^*, \lambda_2^*)$, the inequality constraints (2.34), along with the optimal policy (3.5), need to be evaluated at the constraint boundaries.

$$\begin{pmatrix} -I_m \\ I_m \end{pmatrix} H_u \pi^* = -\Delta H(x, u_{bl}) \quad (3.6)$$

Substituting (3.5) into (3.6), yields

$$\begin{pmatrix} -I_m \\ I_m \end{pmatrix} \left( \underbrace{-H_u R_\pi^{-1} H_u^T}_{R_\lambda} (\lambda_1 - \lambda_2) \right) = 2 \Delta H(x, u_{bl}) \quad (3.7)$$

At this point, the positive definite symmetric weight matrix $R_\lambda$ in (3.7) is forced to become identity, by a proper selection of the cost weight $R_\pi$,

$$R_\pi = H_u^T H_u \quad (3.8)$$

resulting in

$$R_\lambda = H_u R_\pi^{-1} H_u^T = I_m \quad (3.9)$$

and completely decoupling the system of equations (3.7).

$$\begin{pmatrix} I_m \\ -I_m \end{pmatrix} (\lambda_1 - \lambda_2) = 2 \Delta H(x, u_{bl}) \quad (3.10)$$

Based on the complementary slackness conditions from (3.3), the system (3.10) can be decomposed into the two $m$-dimensional mutually exclusive decoupled subsets of scalar equations, each one representing a specific active vector-boundary condition, with at most one nonnegative Lagrange coefficient.

$$\begin{bmatrix} H_1(x, u_{bl}) = 0 \end{bmatrix} \Rightarrow \begin{bmatrix} \boxed{\lambda_1 \geq 0}, & \lambda_2 = 0 \end{bmatrix} \Rightarrow \begin{bmatrix} \lambda_1 = 2 \Delta H_1(x, u_{bl}) \end{bmatrix}$$
$$\begin{bmatrix} H_2(x, u_{bl}) = 0 \end{bmatrix} \Rightarrow \begin{bmatrix} \lambda_1 = 0, & \boxed{\lambda_2 \geq 0} \end{bmatrix} \Rightarrow \begin{bmatrix} \lambda_2 = 2 \Delta H_2(x, u_{bl}) \end{bmatrix} \quad (3.11)$$

Enforcing dual feasibility requirement (3.3), gives the two optimal Lagrange vector coefficients,

$$\lambda_1^* = 2 \max(0_{m \times 1}, \Delta H_1(x, u_{bl}))$$
$$\lambda_2^* = 2 \max(0_{m \times 1}, \Delta H_2(x, u_{bl})) \quad (3.12)$$

and after substituting (3.12) into (3.5), the corresponding min-norm optimal control augmentation policy can be written explicitly.

$$\begin{aligned}
\pi^* &= \underbrace{R_\pi^{-1} H_u^T}_{(H_u^T H_u)^{-1} H_u^T = H_u^{-1}} \left( \max\left(0_{m\times 1}, \Delta H_1(x, u_{bl})\right) - \max\left(0_{m\times 1}, \Delta H_2(x, u_{bl})\right) \right) \\
&= H_u^{-1} \left( \max\left(0_{m\times 1}, \Delta H_1(x, u_{bl})\right) - \max\left(0_{m\times 1}, \Delta H_2(x, u_{bl})\right) \right) \quad (3.13) \\
&= H_u^{-1} \begin{cases} \underbrace{\left(-H_x x - H_u u_{bl} + \alpha_\pi y_{\lim}^{\min}\right)}_{\Delta H_1(x, u_{bl})}, & \text{if } \Delta H_1(x, u_{bl}) > 0 \\ -\underbrace{\left(H_x x + H_u u_{bl} - \alpha_\pi y_{\lim}^{\max}\right)}_{\Delta H_2(x, u_{bl})}, & \text{if } \Delta H_2(x, u_{bl}) > 0 \\ 0, & \text{otherwise} \end{cases}
\end{aligned}$$

**Remark 1.** For analysis and implementation purposes, the component-wise max functions in (3.13) can be replaced by an equivalent algebraic equation,

$$f_{\max} = \max(f) = \frac{f + |f|}{2} \quad (3.14)$$

to transform a scalar argument $f$ into its respective scalar max-bounded output $f_{\max}$. This expression shows that the QP solution (3.13) represents a piece-wise continuous linear state-feedback augmentation controller.

$$\pi^*(x, u_{bl}) = 0.5 H_u^{-1} \left( \Delta H_1(x, u_{bl}) + |\Delta H_1(x, u_{bl})| - \Delta H_2(x, u_{bl}) - |\Delta H_2(x, u_{bl})| \right) \quad (3.15)$$

**Remark 2.** The developed control augmentation design is directly related to the method of Control Barrier Functions (CBF-s) [8], [9]. The two designs use the same three theoretical pillars: 1) The Nagumo theorem, 2) The Comparison lemma, and 3) The Min-norm optimal control design concept. However, the derived augmentation design is very different. First, the "CBF safe" set within the output operational limits does not have to be bounded. Second, differentiability of the set boundary curves is not required and the set does not have to be convex. Finally, CBF conditions are not required to be verified prior to the design. The vector functions that define the operational set boundaries become CBF-s by the design and that is the main difference between the developed and the original CBF-based methods. Nonetheless, in order to acknowledge historical precedence and originality of the CBF design, the developed method shall be often referred to as the "CBF augmentation".

**Remark 3.** The total control signal (1.4) with the CBF analytical solution policy (3.13) can be written as,

$$u = u_{bl} + \pi^* = u_{bl} + H_u^{-1} \begin{cases} \underbrace{\left(-H_x x - H_u u_{bl} + \alpha_\pi y_{\lim}^{\min}\right)}_{\Delta H_1(x, u_{bl})}, & \text{if } \Delta H_1(x, u_{bl}) > 0 \\ -\underbrace{\left(H_x x + H_u u_{bl} - \alpha_\pi y_{\lim}^{\max}\right)}_{\Delta H_2(x, u_{bl})}, & \text{if } \Delta H_2(x, u_{bl}) > 0 \\ 0, & \text{otherwise} \end{cases} \quad (3.16)$$

and it reveals "cancelation" of the baseline controller by the CBF component. For example, if $\Delta H_1(x, u_{bl}) > 0$ then the baseline control component $u_{bl}$ is cancelled and replaced by the CBF augmentation feedback with a feedforward term on the corresponding limit value.

$$u = u_{bl} + H_u^{-1}\left(-H_x x - H_u u_{bl} + \alpha_\pi y_{\lim}^{\min}\right) = H_u^{-1}\left(-H_x x + \alpha_\pi y_{\lim}^{\min}\right) \quad (3.17)$$

For presentation clarity, the derived CBF augmentation design is summarized in the table below.

| Open-loop LTI MIMO plant dynamics (1.1) | $\dot{x} = Ax + Bu$ |
|---|---|
| Limited output (1.1) | $y_{\lim} = C_{\lim} x$ |
| Control input (1.4) | $u = u_{bl} + \pi$ |
| Closed-loop system (1.5) | $\dot{x} = Ax + B(u_{bl} + \pi)$ |
| Limited output constraints (2.2) | $h(x) = \begin{pmatrix} h_1(x) \\ h_2(x) \end{pmatrix} = \begin{pmatrix} y_{\lim}^{\min} - y_{\lim} \\ y_{\lim} - y_{\lim}^{\max} \end{pmatrix} \leq 0$ |
| Stable polynomial for limited output components with relative degree greater than zero (2.6) | $\phi_i(s) = \prod_{j=1}^{r_i}(s - \lambda_{ij}) = \sum_{j=0}^{r_i} c_{ij} s^j, \quad \lambda_{ij} \in \mathbb{R}^-, \quad r_i \geq 1, \quad 1 \leq i \leq m$ |
| Diagonal matrix of stable polynomials (2.8) | $\Phi(s) = \mathrm{diag}(\phi_1(s) \ \ldots \ \phi_m(s))$ |
| Modified limited output and constraints (2.8) | $Y_{\lim} = \Phi(s) y_{\lim}, \quad H(x, u_{bl}, \pi) = \begin{pmatrix} \Phi(s) h_1(x) \\ \Phi(s) h_2(x) \end{pmatrix} \leq 0$ |
| Modified limited output and modified output constraints representation (2.9) | $Y_{\lim} = H_x x + H_u(u_{bl} + \pi)$ <br> $\underbrace{H(x, u_{bl}, \pi)}_{\binom{H_1(x, u_{bl}, \pi)}{H_2(x, u_{bl}, \pi)}} = \begin{pmatrix} -Y_{\lim} + \alpha_\pi y_{\lim}^{\min} \\ Y_{\lim} - \alpha_\pi y_{\lim}^{\max} \end{pmatrix} = \begin{pmatrix} -I_m \\ I_m \end{pmatrix} H_u \pi + \underbrace{\begin{pmatrix} -H_x x - H_u u_{bl} + \alpha_\pi y_{\lim}^{\min} \\ H_x x + H_u u_{bl} - \alpha_\pi y_{\lim}^{\max} \end{pmatrix}}_{\Delta H(x, u_{bl}) = \binom{\Delta H_1(x, u_{bl})}{\Delta H_2(x, u_{bl})}} \leq 0$ |
| Auxiliary data (2.10), (2.11) | $H_x = \begin{pmatrix} (C_{\lim})_1 \phi_1(A) \\ \vdots \\ (C_{\lim})_m \phi_m(A) \end{pmatrix} = \begin{pmatrix} (C_{\lim})_1 \prod_{j=1}^{r_1}(A - \lambda_{1j} I_n) \\ \vdots \\ (C_{\lim})_m \prod_{j=1}^{r_m}(A - \lambda_{mj} I_n) \end{pmatrix}, \quad H_u = \begin{pmatrix} (C_{\lim})_1 A^{r_1-1} B \\ (C_{\lim})_2 A^{r_2-1} B \\ \vdots \\ (C_{\lim})_m A^{r_m-1} B \end{pmatrix}$ <br> $\alpha_\pi = \begin{pmatrix} c_{10} & \ldots & 0 \\ \vdots & \ddots & \vdots \\ 0 & \ldots & c_{m0} \end{pmatrix} > 0, \quad c_{i0} = \prod_{j=1}^{r_i}(-\lambda_{ij}) > 0, \quad \forall i = 1, \ldots, m$ |
| Min-norm optimal CBF control augmentation solution (3.13), (3.15) | $\pi^* = H_u^{-1} \begin{cases} \underbrace{(-H_x x - H_u u_{bl} + \alpha_\pi y_{\lim}^{\min})}_{\Delta H_1(x, u_{bl})}, & \text{if } \Delta H_1(x, u_{bl}) > 0 \\ -\underbrace{(H_x x + H_u u_{bl} - \alpha_\pi y_{\lim}^{\max})}_{\Delta H_2(x, u_{bl})}, & \text{if } \Delta H_2(x, u_{bl}) > 0 \\ 0, & \text{otherwise} \end{cases}$ <br> $= H_u^{-1} \left( \max(0_{m \times 1}, \Delta H_1(x, u_{bl})) - \max(0_{m \times 1}, \Delta H_2(x, u_{bl})) \right)$ <br> $= 0.5 H_u^{-1} \left( \Delta H_1(x, u_{bl}) + |\Delta H_1(x, u_{bl})| - \Delta H_2(x, u_{bl}) - |\Delta H_2(x, u_{bl})| \right)$ |
| Total Control (3.16) | $u = u_{bl} + \pi^* = u_{bl} + H_u^{-1} \begin{cases} \underbrace{(-H_x x - H_u u_{bl} + \alpha_\pi y_{\lim}^{\min})}_{\Delta H_1(x, u_{bl})}, & \text{if } \Delta H_1(x, u_{bl}) > 0 \\ -\underbrace{(H_x x + H_u u_{bl} - \alpha_\pi y_{\lim}^{\max})}_{\Delta H_2(x, u_{bl})}, & \text{if } \Delta H_2(x, u_{bl}) > 0 \\ 0, & \text{otherwise} \end{cases}$ |

**Table 1  Min-Norm Optimal CBF Augmentation Design Summary**

Figure 1 shows the system block-diagram with a baseline state feedback controller augmented by the min-norm optimal state feedback CBF logic for enforcing the desired operational constraints on the system output.

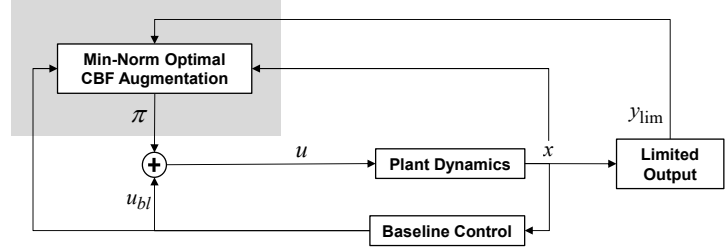

**Figure 1** Closed-loop system block-diagram with min-norm optimal CBF augmentation

By design, the augmentation logic enforces soft constraints on the selected limited output, via feedback and without an explicit hard saturation logic. Also, the derived min-norm optimal control augmentation solution (3.13) represents a continuous piece-wise linear state feedback control policy [8], [9] and as such, the corresponding closed-loop system stability and robustness properties can be directly analyzed using standard methods from linear systems [3].

## 4. Closed-Loop System Dynamics and Uniform Boundedness Analysis

With the optimal CBF policy $\pi^*$ (3.13), the modified limited output $Y_{\lim}$ from (2.9) can be written as,

$$Y_{\lim} = H_x x + H_u \underbrace{\left(u_{bl} + \pi^*\right)}_{u} \tag{4.1}$$

and therefore, the total control can be written as,

$$u = H_u^{-1}\left(Y_{\lim} - H_x x\right) \tag{4.2}$$

Substituting (4.2) into (1.1), gives the closed-loop system dynamics,

$$\dot{x} = \underbrace{\left(A - B H_u^{-1} H_x\right)}_{\tilde{A}_{cl}} x + B H_u^{-1} Y_{\lim} = \tilde{A}_{cl} x + B H_u^{-1} Y_{\lim} \tag{4.3}$$

with the system closed-loop matrix $\tilde{A}_{cl}(x)$,

$$\tilde{A}_{cl} = A - B \underbrace{\left(H_u^{-1} H_x\right)}_{K_{CBF}} \tag{4.4}$$

where

$$K_{CBF} = H_u^{-1} H_x \tag{4.5}$$

represents the CBF augmentation state feedback gain matrix.

Expression (4.3) is essential in proving ultimate boundedness of the closed-loop system trajectories, driven by the component-wise limited modified $m$–dimensional output signal $Y_{\lim}$ (2.15), which in turn is FI by the CBF augmentation design, with respect to the modified $(2m)$–dimensional constraints from (2.9),

$$\begin{pmatrix} -Y_{\lim} + \alpha_\pi y_{\lim}^{\min} \\ Y_{\lim} - \alpha_\pi y_{\lim}^{\max} \end{pmatrix} \leq 0 \tag{4.6}$$

or equivalently,

$$\underbrace{\left(\alpha_\pi y_{\lim}^{\min}\right)}_{Y_{\lim}^{\min}} \leq Y_{\lim} \leq \underbrace{\left(\alpha_\pi y_{\lim}^{\max}\right)}_{Y_{\lim}^{\max}} \tag{4.7}$$

where the above double-inequality is understood component-wise. In other words, trajectories of (4.3) are uniformly bounded forward in time if and only if $\tilde{A}_{cl}$ (4.4) is Hurwitz, whereby the sufficiency is obvious and the necessity can be proven by the contradiction argument.

The iff condition for $\tilde{A}_{cl}$ (4.4) to be Hurwitz for keeping trajectories uniformly bounded also reveals the "cancellation" effect of the CBF augmentation design in the sense that the system closed-loop stability is independent of the baseline controller and it is defined only by the CBF feedback $\left(-H_u^{-1}H_x x\right)$ in (4.3).

The next statement summarizes all of the derived CBF control augmentation properties.

**Theorem 1**. For the stabilizable LTI system (1.1), with the baseline stabilizing state feedback controller $u_{bl}$ from (1.4), the min-norm optimal CBF control augmentation policy $\pi^*$ (3.13) represents a piece-wise linear continuous state feedback, which is designed to enforce the designated operational min/max output constrains (1.2) component-wise. The corresponding output constraint set in (2.34) is forward invariant with respect to the closed-loop system dynamics (4.3). In addition, all of the closed-loop system trajectories with an initial condition $x(t_0) = x_0$, starting at a given time $t_0$ from the operational set, as defined by the min/max box constraints in (2.34), will remain uniformly bounded forward in time, for all $t \geq t_0$ if and only if the closed-loop matrix $\tilde{A}_{cl}$ (4.4) is Hurwitz.

## 5. Relative Stability and Margins Analysis

In addition to closed-loop stability, practical control systems are required to possess relative stability, which is defined in terms of the system gain and phase margins [1], [3], computed at the system input and/or output breakpoints.

Towards that end, consider the total control (4.2),

$$u = H_u^{-1}\left(Y_{\lim} - H_x x\right) = -\underbrace{\left(H_u^{-1}H_x\right)}_{K_{CBF}} x + H_u^{-1} Y_{\lim} = \underbrace{-K_{CBF} x}_{\text{CBF Feedback}} + \underbrace{H_u^{-1} Y_{\lim}}_{\text{Uniformly Bounded Signal}} \quad (5.1)$$

with the CBF feedback gain $K_{CBF}$ (4.5) and a uniformly bounded term $Y_{\lim}(t)$ (4.7). As discussed, the closed-loop system stability is defined by the CBF feedback only and consequently, the system stability margins are defined by the related loop gain transfer function $L_u(s)$, which is independent of $Y_{\lim}(t)$.

$$u_{out} = -K_{CBF} x = -\underbrace{\left[K_{CBF}\left(sI_n - A\right)^{-1} B\right]}_{L_u(s)} u_{in} = -L_u(s) u_{in} \quad (5.2)$$

In other words, gain and time-delay uncertainties in the bounded signal $Y_{\lim}$ do not change the closed-loop system stability.

There is an alternative method to analyzing relative stability. Define a diagonal positive-definite state-dependent matrix,

$$\delta(x) = \begin{pmatrix} \delta_1(x) & \cdots & 0 \\ \vdots & \ddots & \vdots \\ 0 & \cdots & \delta_m(x) \end{pmatrix} \in R^{m \times m} \quad (5.3)$$

with state-dependent non-negative binary-valued diagonal elements,

$$\delta_i(x) = \begin{cases} 1, & \text{if } \left[\left(-H_x x - H_u u_{bl} + \alpha_\pi y_{\lim}^{\min}\right)_i > 0\right] \vee \left[\left(H_x x + H_u u_{bl} - \alpha_\pi y_{\lim}^{\max}\right)_i > 0\right] \\ 0, & \text{otherwise} \end{cases} \quad (5.4)$$

$\forall i = 1, \ldots, m$, which in turn represents a generalization of the continuous switching logic in (3.13). By definition, $\|\delta(x)\| \leq 1$, uniformly in $x$. Then the total control can be written as,

$$\begin{aligned} u &= u_{bl} + \pi^* = u_{bl} - H_u^{-1}\delta(x)\left(H_x x + H_u u_{bl} - \alpha_\pi y_{\lim}^{\min/\max}\right) \\ &= \underbrace{\left(I_m - H_u^{-1}\delta(x)H_u\right)u_{bl}}_{\text{Scaled Baseline Control}} - \underbrace{H_u^{-1}\delta(x)H_x x}_{\text{CBF Feedback}} + \underbrace{H_u^{-1}\delta(x)\alpha_\pi y_{\lim}^{\min/\max}}_{\text{CBF Command}} \end{aligned} \quad (5.5)$$

where the $i^{th}$ component of the constant vector $y_{\lim}^{\min/\max} \in R^m$ is defined below.

$$\left(y_{\lim}^{\min/\max}\right)_i = \begin{cases} \left(y_{\lim}^{\min}\right)_i, & \text{if } \left(-H_x x - H_u u_{bl} + \alpha_\pi y_{\lim}^{\min}\right)_i > 0 \\ \left(y_{\lim}^{\max}\right)_i, & \text{if } \left(H_x x + H_u u_{bl} - \alpha_\pi y_{\lim}^{\max}\right)_i > 0 \\ 0, & \text{otherwise} \end{cases} \quad (5.6)$$

The CBF controller (5.5) remains continuous since the switching function $\delta(x)$ is multiplied by the condition-dependent continuous functions from (5.4), (5.6).

**Remark 4.** It is interesting to note that the CBF augmentation signal (5.5) can be viewed as a continuous piece-wise linear state feedback linearizing controller [10], [12], with the embedded switching logic (5.4). In other words, (5.5) represents a dynamic inversion (DI) controller. The DI nature of (5.5) is due to the feedback gain $\left(H_u^{-1}\delta(x)H_x\right)$ and because of that, the corresponding closed-loop system properties can also be analyzed within the DI framework.

**Remark 5.** Suppose that the baseline controller is designed in the state-proportional feedback form,

$$u_{bl} = -K_x x \quad (5.7)$$

with the feedback gain $K_x \in R^{m \times n}$ selected to make the corresponding closed-loop system matrix Hurwitz. Then (5.5) becomes,

$$\begin{aligned} u &= \left(I_m - H_u^{-1}\delta(x)H_u\right)(-K_x x) - H_u^{-1}\delta(x)H_x x + H_u^{-1}\delta(x)\alpha_\pi y_{\lim}^{\min/\max} \\ &= -\left(\left(I_m - H_u^{-1}\delta(x)H_u\right)K_x + H_u^{-1}\delta(x)H_x\right)x + H_u^{-1}\delta(x)\alpha_\pi y_{\lim}^{\min/\max} \end{aligned} \quad (5.8)$$

Substituting (5.8) into the open-loop dynamics (1.1), gives the closed-loop system.

$$\dot{x} = Ax - B\left(\left(I_m - H_u^{-1}\delta(x)H_u\right)K_x + H_u^{-1}\delta(x)H_x\right)x + BH_u^{-1}\delta(x)\alpha_\pi y_{\lim}^{\min/\max} \quad (5.9)$$

For relative stability analysis, it is assumed that $\delta(x)$ (5.3) is a constant diagonal matrix, with binary values on its diagonal. Also, only the feedback portion of the CBF controller (5.5) needs to be considered and in that case, relative stability metrics, such as gain and phase margins, can be computed and analyzed based on the overall system block-diagram, shown in Figure 2.

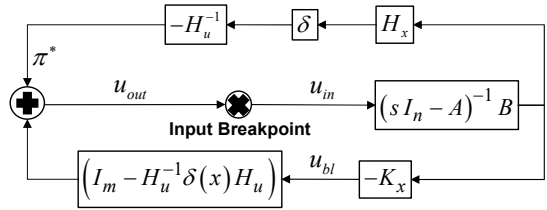

**Figure 2 Relative stability analysis of the closed-loop system with CBF augmentation**

Similar to the previous method for relative stability analysis, in this case stability margins are defined by the $(m \times m)$ system loop gain $L_u(s)$ transfer function matrix, computed at the control input break point. It describes signal propagation dynamics from $u_{in}$ to $u_{out}$.

$$u_{out} = -\underbrace{\left[\left(\left(I_m - H_u^{-1}\delta(x)H_u\right)K_x + H_u^{-1}\delta(x)H_x\right)(sI_n - A)^{-1}B\right]}_{L_u(s;\delta)} u_{in} = -L_u(s;\delta)u_{in} \qquad (5.10)$$

Using the system loop gain from (5.10),

$$L_u(s;\delta) = \left(\left(I_m - H_u^{-1}\delta(x)H_u\right)K_x + H_u^{-1}\delta(x)H_x\right)(sI_n - A)^{-1}B \qquad (5.11)$$

parameterized with $\delta$, SISO and MIMO gain / phase margins can be computed [3] for all possible combinations of the binary-valued diagonal elements of $\delta$.

## 6. CBF Augmentation Design for State Feedback Servo-Controllers with Command-Feedforward Connections

Often in practice, it is of interest to design servo-controllers that are able to track external bounded commands. In that case, the baseline servo-control policy can be represented as a sum of two terms, a proportional state feedback $u_P$ and a command-feedforward $u_{ff}$.

$$u_{bl} = \underbrace{-K_x x}_{u_P} + \underbrace{K_{ff} y_{cmd}}_{u_{ff}} = u_P + u_{ff} \qquad (6.1)$$

In (6.1), the feedforward gain matrix $K_{ff} \in R^{m \times m}$ can be computed such that the DC gain from the command to the system regulated output is unity.

$$K_{ff} = \begin{pmatrix} K_x & I_m \end{pmatrix} \begin{pmatrix} A & B \\ C_{reg} & D_{reg} \end{pmatrix}^{-1} \begin{pmatrix} 0_{n \times m} \\ I_m \end{pmatrix} \qquad (6.2)$$

Note that the matrix invertibility in (6.2) is a standard assumption for the design of servo-controllers [3]. It implies that there are no transmission zeros at the origin in the system input-to-output dynamics.

The total control signal is defined as in (1.4),

$$u = u_{bl} + \pi \qquad (6.3)$$

resulting in the corresponding closed-loop system dynamics (1.5).

$$\dot{x} = Ax + B\underbrace{(u_{bl} + \pi)}_{u} \qquad (6.4)$$

For the system (6.4), the CBF augmentation design goal remains the same: Find the min-optimal state feedback policy $\pi^*(x)$ such that the output constraints (1.2) are enforced.

In this case, the min-norm optimal control solution $\pi^*$ (3.13) is defined as an augmentation to the baseline controller $u_{bl}$ (6.1).

$$\pi^* = H_u^{-1} \begin{cases} \underbrace{\left(-H_x x - H_u u_{bl} + \alpha_\pi y_{\lim}^{\min}\right)}_{\Delta H_1(x,\, u_{bl})}, & \text{if } \Delta H_1(x, u_{bl}) > 0 \\ -\underbrace{\left(H_x x + H_u u_{bl} - \alpha_\pi y_{\lim}^{\max}\right)}_{\Delta H_2(x,\, u_{bl})}, & \text{if } \Delta H_2(x, u_{bl}) > 0 \\ 0, & \text{otherwise} \end{cases}$$

$$= H_u^{-1} \begin{cases} \underbrace{\left(-(H_x + H_u K_x)x - H_u K_{ff} y_{cmd} + \alpha_\pi y_{\lim}^{\min}\right)}_{\Delta H_1}, & \text{if } \Delta H_1 > 0 \\ -\underbrace{\left((H_x + H_u K_x)x + H_u K_{ff} y_{cmd} - \alpha_\pi y_{\lim}^{\max}\right)}_{\Delta H_2}, & \text{if } \Delta H_2 > 0 \\ 0, & \text{otherwise} \end{cases} \qquad (6.5)$$

Closed-loop system block-diagram.

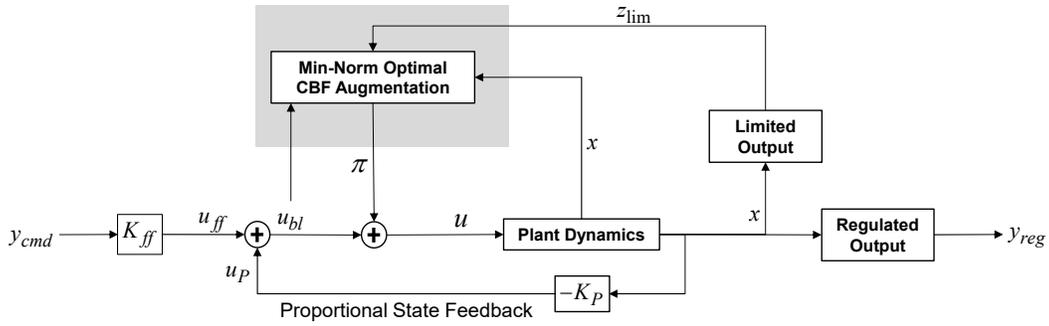

**Figure 3  Closed-loop system with baseline proportional servo-controller and CBF augmentation**

Based on (6.5), closed-loop system stability needs to be verified only if and when the system trajectories evolve on their respective bounds. Otherwise, $\pi = 0$ and the closed-loop system operates under the baseline servo-controller (6.1).

## 7. CBF Augmentation Design for Scalar LTI Dynamics

Consider the CBF augmentation design for the scalar LTI dynamics.

$$\dot{x} = a x + b u \qquad (7.1)$$

The CBF design goal is to augment the baseline servo-controller,

$$u_{bl} = -k_x x + k_{ff} x_{cmd} \qquad (7.2)$$

with a CBF state feedback,

$$\pi = \pi(x) \qquad (7.3)$$

to enforce the system state min/max limits,

$$\underbrace{x_{\min}}_{y_{\lim}^{\min}} \leq \underbrace{x}_{y_{\lim}} \leq \underbrace{x_{\max}}_{y_{\lim}^{\max}} \qquad (7.4)$$

for all trajectories of the resulting closed-loop system,

$$\dot{x} = (a - b k_x) x + b(k_{ff} x_{cmd} + \pi) \qquad (7.5)$$

where the baseline feedback gain $k_x$ is selected such that $(a - b k_x) < 0$, the feedforward gain $k_{ff}$ is defined such that the system state $x$ tracks an external bounded command $x_{cmd}$ with zero steady state errors.

$$k_{ff} = (k_x \quad 1) \begin{pmatrix} a & b \\ 1 & 0 \end{pmatrix}^{-1} \begin{pmatrix} 0 \\ 1 \end{pmatrix} \tag{7.6}$$

The limited output relative degree is unity. Let $\lambda < 0$ be the eigenvalue of the corresponding stable polynomial (2.6). Then the modified output $Y_{\lim}$ is defined by (2.9),

$$Y_{\lim} = \dot{x} - \lambda x = \underbrace{(a-\lambda)}_{H_x} x + \underbrace{b}_{H_u} (u_{bl} + \pi) \tag{7.7}$$

and the CBF augmentation feedback is in the form of (3.13), with $\alpha_\pi = -\lambda$.

$$\pi^* = H_u^{-1} \begin{cases} \underbrace{\left(-H_x x - H_u u_{bl} + \alpha_\pi y_{\lim}^{\min}\right)}_{\Delta H_1(x, u_{bl})}, & \text{if } \Delta H_1(x, u_{bl}) > 0 \\ \underbrace{-\left(H_x x + H_u u_{bl} - \alpha_\pi y_{\lim}^{\max}\right)}_{\Delta H_2(x, u_{bl})}, & \text{if } \Delta H_2(x, u_{bl}) > 0 \\ 0, & \text{otherwise} \end{cases} \tag{7.8}$$

$$= b^{-1} \begin{cases} \left(-(a - b k_x - \lambda) x - b k_{ff} x_{cmd} - \lambda x_{\min}\right), & \text{if } \Delta H_1 > 0 \\ \left(-(a - b k_x - \lambda) x - b k_{ff} x_{cmd} + \lambda x_{\max}\right), & \text{if } \Delta H_2 > 0 \\ 0, & \text{otherwise} \end{cases}$$

Consider the first logical path in (7.8). It becomes active if the first modified constraint increment,

$$\Delta H_1 = -\underbrace{(a - b k_x) x - b k_{ff} x_{cmd}}_{\dot{x}|_{u=u_{bl}}} + \lambda (x - x_{\min}) \tag{7.9}$$

is violated under the baseline controller.

In that case, the CBF controller cancels the baseline servo-controller dynamics and forces the closed-loop system asymptotically approach the designated minimum boundary of the state limit.

$$\left[\Delta H_1(x) > 0\right] \Rightarrow \left[\dot{x} = \lambda (x - x_{\min})\right] \tag{7.10}$$

Similar arguments apply to analyzing the closed-loop system dynamics when the second logic path in (7.8) becomes active.

$$\left[\Delta H_2(x) > 0\right] \Rightarrow \left[\dot{x} = \lambda (x - x_{\max})\right] \tag{7.11}$$

It is of interest to compare the CBF augmentation controller (7.8) to that of the rectangular Projection Operator [3], which can also be applied to (7.5) as an augmentation logic to keep the closed-loop system trajectories evolving within the prescribed min/max limits (7.4). For two scalar inputs, $x$ and $y$, the Projection Operator control augmentation logic can be written as,

$$\text{Proj}(x, y) = \begin{cases} \left(\dfrac{x_{\max} - x}{\delta}\right) y, & \left[(x > x_{\max} - \delta) \wedge (y > 0)\right] \\ \left(\dfrac{x - x_{\min}}{\delta}\right) y, & \left[(x < x_{\min} + \delta) \wedge (y < 0)\right] \\ y, & \text{otherwise} \end{cases} \tag{7.12}$$

where $\delta > 0$ is a sufficiently small positive constant. Using the system dynamics (7.5), gives

$$\dot{x} = \underbrace{\left[(a - bk_x)x + bk_{ff} x_{cmd}\right]}_{\dot{x}_{bl}} + b\pi_{proj}(x) = \text{Proj}\left(x, \underbrace{y}_{\dot{x}_{bl}}\right) \quad (7.13)$$

where $\pi_{proj}(x)$ denotes the Projection Operator feedback,

$$\pi_{proj}(x) = b^{-1}\left(\text{Proj}(x, \dot{x}|_{bl}) - \dot{x}|_{bl}\right) = b^{-1}\begin{cases} \left(\dfrac{x_{max} - x}{\delta} - 1\right)\dot{x}|_{bl}, & \left[(x > x_{max} - \delta) \wedge (\dot{x}|_{bl} > 0)\right] \\ \left(\dfrac{x - x_{min}}{\delta} - 1\right)\dot{x}|_{bl}, & \left[(x < x_{min} + \delta) \wedge (\dot{x}|_{bl} < 0)\right] \\ 0 & , \text{ otherwise} \end{cases} \quad (7.14)$$

which in turn represent a nonlinear state feedback controller. So, the CBF linear feedback controller (7.8) keeps the LTI property of the system, while the Projection Operator based nonlinear control logic (7.14) does not.

Figure 4 shows simulation comparison data for $a = b = c = 1$, $k_x = 4$ and $k_{ff} = 3$.

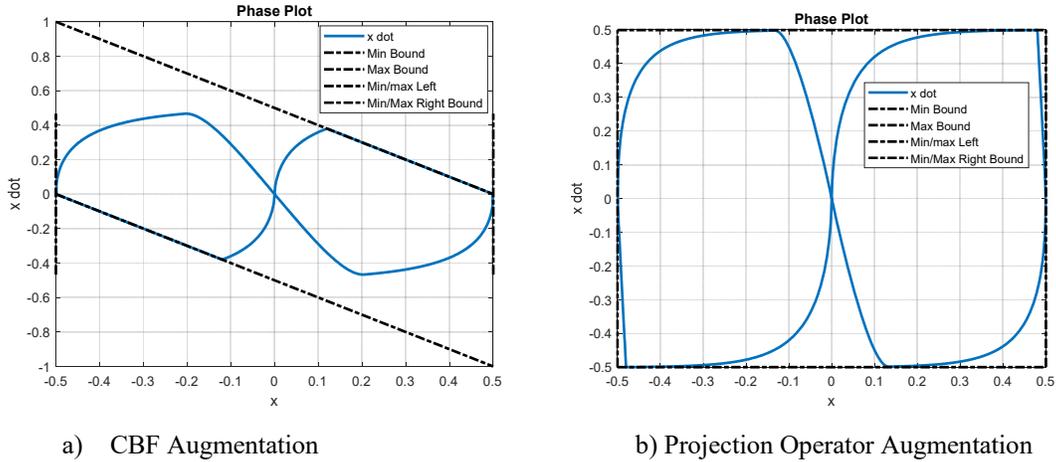

a)   CBF Augmentation                                                    b) Projection Operator Augmentation

**Figure 4  CBF and Projection Operator Augmentation Design Comparison: Phase Data**

CBF augmentation controller (7.8) is designed with $\lambda = -1$. Tolerance for the projection operator controller (7.14) is $\delta = 0.01$. The target min/max state limits are selected the same for both designs, $x_{min} = -0.5$ and $x_{max} = 0.5$. The simulation data show similar dynamics between the two cases, with a slightly more conservative achievable rate for the CBF controller. Total control input data comparison is presented in Figure 5.

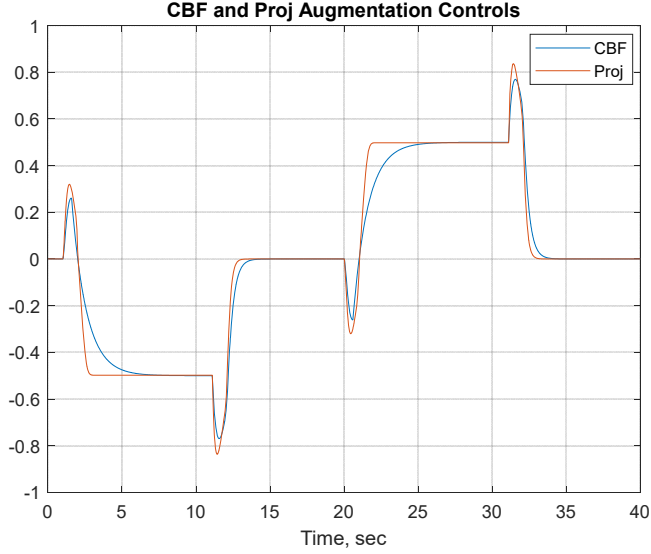

**Figure 5 CBF and Projection Operator Augmentation Design Comparison: Total Control Data**

The data are very similar. However, the Projection Operator controller is more aggressive in enforcing the desired min/max limits on the system state.

In general, there is no clear "winner" in this test case. Both controllers yield adequate closed-loop system time-domain performance, with the imposed box limits on the system state. However, margins of the CBF controller can be analyzed, while the inherent nonlinearity of the Projection Operator controller prevents an analysis of the corresponding system relative stability metrics.

## 8. Dynamic Servo-Control Augmentation Design with Input and Output Constraints

In the previous sections, the CBF augmentation state feedback solution is presented for MIMO LTI systems with proportional state feedback baseline controllers to address stabilization and servo-control problems with component-wise output constraints. In most if not all practical applications, of primary interest is the design of Proportional-Integral (PI) state feedback servo-controllers for command tracking. This section extends the results of Theorem 1 to the design of PI state feedback servo-controllers with CBF augmentation to enforce input and output min/max operational limits.

### A. Constrained Servo-Control Problem Formulation and Solution

Consider the controllable LTI MIMO system,

$$\text{Open-Loop System Dynamics}: \dot{x}_p = A_p x_p + B_p u, \quad x_p \in R^{n_p}$$
$$\text{Regulated Output}: y_{reg} = C_{p\,reg} x_p + D_{p\,reg} u, \quad y_{reg} \in R^m \quad (8.1)$$
$$\text{Limited Output}: z_{\lim} = C_{p\lim} x_p, \quad z_{\lim} \in R^m$$

where $x_p \in R^{n_p}$ is the $n_p$-dimensional state vector, $u \in R^m$ is the $m$-dimensional control input, $y_{reg} \in R^m$ is the system $m$-dimensional vector of regulated outputs, and $z_{\lim} \in R^m$ is the $m$-dimensional limited output to be kept within the desired min/max bounds $\left(z_{\lim}^{\min}, z_{\lim}^{\max}\right) \in R^m$, component-wise.

In (8.1), the system matrices $\left(A_p, B_p, C_{p\lim}, C_{p\,reg}, D_{p\,reg}\right)$ are of the corresponding dimensions and the matrix pair $\left(A_p, B_p\right)$ is controllable. It is further assumed that the system state vector $x_p$ is accessible for control design, as the system output measurement.

In this section, of interest is the CBF augmentation design for servo-controllers with min/max operational constraints on the system baseline control input $u_{bl}$ and on the selected limited output $z_{lim}$. Specifically, a state feedback CBF control augmentation policy $w$ needs to be found such that the total control $u = u_{bl} + w$ forces the system regulated output $y_{reg}$ track external commands $y_{cmd} \in R^m$ to the extent possible, while the baseline control input $u_{bl}$ and the system limited output $z_{lim}$ evolve within their predefined min/max operational constraint bounds, component-wise.

$$u_{bl}^{\min} \leq u_{bl} \leq u_{bl}^{\min}, \quad z_{lim}^{\min} \leq z_{lim} \leq z_{lim}^{\max} \tag{8.2}$$

**Assumption 1, (Constraints Feasibility).** The operational constrains are feasible, that is there exists a control strategy $u$ such that the operational limits (8.2) hold along the system (8.1) trajectories, forward in time.

**Remark 5.** Assumed feasibility of input/output constraints (8.2) becomes critical especially for open-loop unstable systems, where control limits may cause system instability. In other words, for unstable systems a certain amount of control is required for stabilization and on top of that, extra control activity may be needed for command tracking. Preserving closed-loop stability is the essential ingredient for any controller. In practical applications, limits on control would be placed first and after that, the corresponding achievable output constraints would be computed. Alternatively, input limits can be derived based on the desired output constraints. In either case, the box limits (8.2) must be achievable, and thus feasible.

Clearly, if the input and/or the limited output constraint saturations become active, the system command tracking performance may and will degrade. In that case, the closed-loop dynamics must remain stable and the system regulated output shall track the external commands "to the extent possible." This section presents derivations of a state feedback PI servo-controller with soft operational constraints (8.2).

In order to facilitate robust tracking of external commands while operating in the presence of the limits (8.2), consider the integrated output tracking error dynamics with the anti-windup (AW) control modification term $v \in R^m$,

$$\dot{e}_{yI} = \underbrace{\left(y_{reg} - y_{cmd}\right)}_{e_y} + v \tag{8.3}$$

added to the output tracking error signal $e_y$, and to be designed such that the integrator state $e_{yI} \in R^m$ is uniformly bounded during control saturation events [1], [2]. The AW signal $v$ in (8.3) can also be viewed as a command modification logic constructed to prevent divergence of the integrator state if the original command cannot be followed due to the imposed operational limits (8.2).

During control saturation events, the controller integrator state $e_{yI}$ needs to be kept bounded irrespective of the tracking error dynamics, which in turn can drive the integrator state to become unbounded, that is the integrator state would "wind-up" [1], [2]. The AW control modification input $v$ in (8.3) will be designed to prevent the integrator state from winding up.

The total control signal is defined as,

$$u = \underbrace{\left( \underbrace{-K_I \, e_{yI}}_{u_I} \underbrace{-K_P \, x_p}_{u_P} \right)}_{u_{bl}} + w = u_{bl} + w \tag{8.4}$$

where the baseline PI servo-controller,

$$u_{bl} = \underbrace{-K_I \, e_{yI}}_{u_I} \underbrace{-K_P \, x_p}_{u_P} = -\underbrace{\begin{pmatrix} K_I & K_P \end{pmatrix}}_{K_x} \underbrace{\begin{pmatrix} e_{yI} \\ x_p \end{pmatrix}}_{x} = -K_x \, x \tag{8.5}$$

is constructed to force the regulated output $y_{reg}$ follow external bounded commands $y_{cmd} \in R^m$ with a prescribed precision, without any limitations on the system input or selected output.

In (8.5), $K_I \in R^{m \times m}$ and $K_P \in R^{m \times n_p}$ represent the integral and the proportional feedback gain matrices, respectively, while $u_I$ and $u_P$ are the integral and the state-proportional feedback components of the baseline PI servo-controller (8.5).

Baseline PI feedback gains can be computed using control-theoretic design methods, such as the Pole placement or the Linear Quadratic Regulator (LQR) [3]. For the servo-control problem to be well-posed, it is assumed that $\det \begin{pmatrix} A_p & B_p \\ C_{p\,reg} & D_{p\,reg} \end{pmatrix} \neq 0$, which means that the original system (8.1) with the regulated output $y_{reg}$ has no transmission zeros at the origin [3].

Next, verifiable sufficient conditions are stated (Lemma 3) to show that imposing min/max operational constraints on the baseline controller $u_{bl}$ enforces the same limits on the total control input, component-wise.

**Lemma 3.** For the LTI MIMO system (8.1), suppose that a piece-wise linear continuous control augmentation signal $w$ is found such that at any given time $t$, it depends only on $z_{\lim}$ and enforces min/max output limits in (8.2) component-wise, forward in time,

$$w = \begin{cases} F_{\min} z_{\lim}, & \text{if } z_{\lim} > z_{\lim}^{\min} \\ F_{\max} z_{\lim}, & \text{if } z_{\lim} < z_{\lim}^{\max} \\ 0, & \text{otherwise} \end{cases} \Rightarrow \boxed{z_{\lim}^{\min} \leq z_{\lim} \leq z_{\lim}^{\max}} \tag{8.6}$$

where $F_{\min}$ and $F_{\max}$ are linear operators that define $w$ in terms of $z_{\lim}$.

Also suppose that an AW augmentation signal $v$ is designed to keep the baseline control $u_{bl}$ within its designated min/max input limits (8.2) and the corresponding logic depends on $w$ and $u_{bl}$.

$$v = \begin{cases} G_{\max} \begin{pmatrix} w \\ u_{bl} \end{pmatrix}, & \text{if } u_{bl} < u_{bl}^{\max} \\ G_{\min} \begin{pmatrix} w \\ u_{bl} \end{pmatrix}, & \text{if } u_{bl} > u_{bl}^{\min} \\ 0, & \text{otherwise} \end{cases} \Rightarrow \boxed{u_{bl}^{\min} \leq u_{bl} \leq u_{bl}^{\max}} \tag{8.7}$$

where $G_{\min}$ and $G_{\max}$ are linear operators that define $v$ in terms of $w$ and $u_{bl}$.

In addition, assume that the closed-loop system with the baseline controller $u_{bl}$ is stable. Then $v$ and $w$ are uniformly bounded and the total control signal satisfies the same min/max limits as the baseline control component, component-wise.

$$u_{bl}^{\min} \leq \underbrace{(u_{bl} + w)}_{u} \leq u_{bl}^{\max} \tag{8.8}$$

***Proof of Lemma 3***: If it is proven that the total control $u$ is bounded then bondedness of $v$ and $w$ follows directly from the definitions (8.7) and (8.6), respectively. So, the focus is on proving uniform boundedness of the total control input, forward in time.

If $z_{\lim}$ is within its min/max bounds but a component $(u_{bl})_i$ of the baseline signal reaches its min or max value then $w = 0$ and $v \neq 0$ from (8.7) will enforce the total control min/max limits (8.8) forward in time.

Without a loss of generalization, suppose that output bounds are symmetric around the origin, that is $-z_{\lim}^{\min} = z_{\lim}^{\max} > 0$. Since the closed-loop system under the baseline control is assumed to be stable then the corresponding closed-loop input-to-output map is stable in the sense of the $L_\infty$ – norm, [12].

$$\max_{t \geq 0} \|z_{\lim}(t)\|_\infty = \|z_{\lim}\|_{L_\infty} \leq \gamma \|u_{bl}\|_{L_\infty} + \beta \tag{8.9}$$

where $\gamma$ and $\beta$ are positive constants, independent of $u_{bl}$. Suppose that at a time $t_*$ the output reaches its bound, when operating under the baseline control input.

$$\|z_{\lim}(t_*)\|_\infty = z_{\lim}^{\max} \leq \gamma \|u_{bl}(t_*)\|_\infty + \beta \tag{8.10}$$

Then by design, an augmentation policy $w(t_*)$ exists, such that the total control forces the output to be within the designated bounds.

$$\|z_{\lim}(t_*)\|_\infty \leq \gamma \|u_{bl}(t_*) + w(t_*)\|_\infty + \beta \leq z_{\lim}^{\max} \tag{8.11}$$

For example, the inversion-based augmentation solution

$$\boxed{w(t_*) = \frac{1}{\gamma}\left(-u_{bl}(t_*) + \frac{z_{\lim}(t_*)}{\|z_{\lim}(t_*)\|_\infty} z_{\lim}^{\max}\right) - \beta} \Rightarrow \boxed{\|z_{\lim}(t_*)\|_\infty = \gamma \|u_{bl}(t_*) + w(t_*)\|_\infty + \beta = z_{\lim}^{\max}} \tag{8.12}$$

forces $z_{\lim}(t_*)$ into the exact desired bound. Substituting (8.10) into (8.11), gives

$$\gamma \|u_{bl}(t_*) + w(t_*)\|_\infty + \beta \leq \gamma \|u_{bl}(t_*)\|_\infty + \beta \tag{8.13}$$

and therefore,

$$\|u_{bl}(t_*) + w(t_*)\|_\infty \leq \|u_{bl}(t_*)\|_\infty \leq u_{bl}^{\max} \tag{8.14}$$

which implies (8.8). The proof of Lemma 3 is complete. □

Combining the system dynamics (8.1) with the total controller (8.4), results in the extended system, driven by an external bounded command $y_{cmd}$ and the CBF augmentation inputs $(v, w)$,

$$\begin{aligned}\dot{e}_{yI} &= C_{p\,reg} x_p + D_{p\,reg}(u_{bl} + w) + \underbrace{\left(-y_{cmd} + v\right)}_{v_{bl}} \\ \dot{x}_p &= A_p x_p + B_p(u_{bl} + w)\end{aligned} \tag{8.15}$$

where

$$v_{bl} = -y_{cmd} \tag{8.16}$$

represents the external command component of the baseline controller. Equivalently, the extended system dynamics (8.15) can be written in a vector form similar to (1.5),

$$\underbrace{\begin{pmatrix} \dot{e}_{yI} \\ \dot{x}_p \end{pmatrix}}_{\dot{x}} = \underbrace{\begin{pmatrix} 0_{m \times m} & C_{p\,reg} \\ 0_{n_p \times m} & A_p \end{pmatrix}}_{A} \underbrace{\begin{pmatrix} e_{yI} \\ x_p \end{pmatrix}}_{x} + \underbrace{\begin{pmatrix} I_m & D_{p\,reg} \\ 0_{n \times m} & B_p \end{pmatrix}}_{\tilde{B}} \underbrace{\begin{pmatrix} v_{bl} + v \\ u_{bl} + w \end{pmatrix}}_{\tilde{u} = \tilde{u}_{bl} + \pi} \tag{8.17}$$

with the total extended control state feedback policy,

$$\tilde{u} = \begin{pmatrix} v_{bl} + v \\ u_{bl} + w \end{pmatrix} = \underbrace{\begin{pmatrix} v_{bl} \\ u_{bl} \end{pmatrix}}_{\tilde{u}_{bl} \in R^{2m}} + \underbrace{\begin{pmatrix} v \\ w \end{pmatrix}}_{\pi \in R^{2m}} = \tilde{u}_{bl} + \pi \tag{8.18}$$

comprised of the extended baseline controller $\tilde{u}_{bl} \in R^{2m}$, augmented by the CBF signal $\pi \in R^{2m}$, whose main purpose is to enforce the desired operational box constraints (8.2) via a proper selection of the two $m$-dimensional CBF augmentation inputs $v$ and $w$. Note that the CBF augmentation policy definition in (8.18) embeds the external command $y_{cmd}$, as shown in (8.16).

Next, the input / output constraints (8.2) are written in terms of the extended system dynamics (8.17),

$$y_{\lim} = \begin{pmatrix} u_{bl} \\ z_{\lim} \end{pmatrix} = \underbrace{\begin{pmatrix} -K_I & -K_P \\ 0_{m \times m} & C_{p \lim} \end{pmatrix}}_{C_{\lim}} \underbrace{\begin{pmatrix} e_{yI} \\ x_p \end{pmatrix}}_{x} = C_{\lim} x \tag{8.19}$$

which in turn, results in the operational constraints expressed in the form of (1.5).

$$\begin{aligned} \text{Min Constraints}: y_{\lim}^{\min} - C_{\lim} x \leq 0 \\ \text{Max Constraints}: C_{\lim} x - y_{\lim}^{\max} \leq 0 \end{aligned} \tag{8.20}$$

The extended system (8.17) and the limited output constraints are in the form of (1.1). In addition, it can be shown that the system matrix pair $(A, \tilde{B})$ is controllable. Consequently, the CBF augmentation design from Table 1 is directly applicable to analytically solving the corresponding QP (2.34), for the extended system dynamics (8.17), with the operational constraints (8.20).

Using (3.13), gives the min-norm optimal piece-wise linear continuous state feedback CBF augmentation policy,

$$\pi^* = H_{\tilde{u}}^{-1} \begin{cases} \underbrace{\left(-H_x x - H_{\tilde{u}} \tilde{u}_{bl} + \alpha_\pi y_{\lim}^{\min}\right)}_{\Delta H_1(x, \tilde{u}_{bl})}, & \text{if } \Delta H_1(x, \tilde{u}_{bl}) > 0 \\ \underbrace{-\left(H_x x + H_{\tilde{u}} \tilde{u}_{bl} - \alpha_\pi y_{\lim}^{\max}\right)}_{\Delta H_2(x, \tilde{u}_{bl})}, & \text{if } \Delta H_2(x, \tilde{u}_{bl}) > 0 \\ 0, & \text{otherwise} \end{cases} \tag{8.21}$$

with the corresponding CBF augmentation parameters.

$$H_x = \begin{pmatrix} (C_{\lim})_1 \phi_1(A) \\ \vdots \\ (C_{\lim})_{(2m)} \phi_{2m}(A) \end{pmatrix} = \begin{pmatrix} (C_{\lim})_1 \prod_{j=1}^{r_1}(A - \lambda_{1j} I_n) \\ \vdots \\ (C_{\lim})_{(2m)} \prod_{j=1}^{r_m}(A - \lambda_{(2m)j} I_n) \end{pmatrix}, \quad H_{\tilde{u}} = \begin{pmatrix} (C_{\lim})_1 A^{r_1 - 1} \tilde{B} \\ \vdots \\ (C_{\lim})_{(2m)} A^{r_{2m} - 1} \tilde{B} \end{pmatrix}$$

$$\alpha_\pi = \begin{pmatrix} c_{10} & \cdots & 0 \\ \vdots & \ddots & \vdots \\ 0 & \cdots & c_{(2m)0} \end{pmatrix} > 0, \quad c_{i0} = \prod_{j=1}^{r_i}(-\lambda_{ij}) > 0, \quad \forall i = 1, \ldots, 2m \tag{8.22}$$

$$\Delta H(x, \tilde{u}_{bl}) = \begin{pmatrix} \Delta H_1(x, \tilde{u}_{bl}) \\ \Delta H_2(x, \tilde{u}_{bl}) \end{pmatrix} = \begin{pmatrix} -H_x x - H_{\tilde{u}} \tilde{u}_{bl} + \alpha_\pi y_{\lim}^{\min} \\ H_x x + H_{\tilde{u}} \tilde{u}_{bl} - \alpha_\pi y_{\lim}^{\max} \end{pmatrix}$$

Substituting (8.21) into (8.18), yields the total control solution, with the piece-wise continuous min-norm optimal CBF augmentation policy,

$$\tilde{u} = \tilde{u}_{bl} + \pi^* = \tilde{u}_{bl} + H_{\tilde{u}}^{-1} \begin{cases} \underbrace{\left(-H_x x - H_{\tilde{u}} \tilde{u}_{bl} + \alpha_\pi y_{\lim}^{\min}\right)}_{\Delta H_1(x,\tilde{u}_{bl})}, & \text{if } \Delta H_1(x,\tilde{u}_{bl}) > 0 \\ \underbrace{-\left(H_x x + H_{\tilde{u}} \tilde{u}_{bl} - \alpha_\pi y_{\lim}^{\max}\right)}_{\Delta H_2(x,\tilde{u}_{bl})}, & \text{if } \Delta H_2(x,\tilde{u}_{bl}) > 0 \\ 0, & \text{otherwise} \end{cases} \quad (8.23)$$

defined component-wise.

### B. Limited Output Vector Relative Degree Verification

Before proceeding any further, the inverse of $H_{\tilde{u}}$ needs to be justified and that leads to verification of the extended system output relative degree. Specifically in this case, the CBF augmentation design (8.21), (8.22) depends on the vector relative degree of the system limited output $y_{\lim}$, with respect to the total control input $\tilde{u}$ (8.18). Using (8.22), consider the modified limited output signal.

$$Y_{\lim} = \begin{pmatrix} U_{bl} \\ Z_{\lim} \end{pmatrix} = \begin{pmatrix} \phi_1(s) u_{bl} \\ \vdots \\ \phi_m(s) u_{bl} \\ \phi_{m+1}(s) z_{\lim} \\ \vdots \\ \phi_{(2m)}(s) z_{\lim} \end{pmatrix} = \underbrace{\begin{pmatrix} \phi_1(s) & \cdots & 0 \\ \vdots & \ddots & \vdots \\ 0 & \cdots & \phi_{(2m)}(s) \end{pmatrix}}_{\Phi(s)} \underbrace{\begin{pmatrix} u_{bl} \\ z_{\lim} \end{pmatrix}}_{y_{\lim}}$$

$$= \underbrace{\begin{pmatrix} (H_{\tilde{u}})_{1,1} & (H_{\tilde{u}})_{1,2} \\ (H_{\tilde{u}})_{2,1} & (H_{\tilde{u}})_{2,2} \end{pmatrix}}_{H_{\tilde{u}}} \underbrace{\begin{pmatrix} v_{bl} + v \\ u_{bl} + w \end{pmatrix}}_{\tilde{u}} + \underbrace{\begin{pmatrix} (H_x)_{1,1} & (H_x)_{1,2} \\ (H_x)_{2,1} & (H_x)_{2,2} \end{pmatrix}}_{H_x} \underbrace{\begin{pmatrix} e_{yI} \\ x_p \end{pmatrix}}_{x} = H_{\tilde{u}} \tilde{u} + H_x x \quad (8.24)$$

where the sensitivity matrices $(H_{\tilde{u}}, H_x)$ are partitioned to show their direct dependence on the total extended control $\tilde{u}$ and on the extended system state components $(e_{yI}, x_p)$. The sensitivity matrix $H_{\tilde{u}}$ is required to be nonsingular and that is verified next.

By definition, the modified baseline control input has vector relative degree one.

$$U_{bl} = \begin{pmatrix} \underbrace{(s-\lambda_1)}_{\phi_1(s)} & \cdots & 0 \\ \vdots & \ddots & \vdots \\ 0 & \cdots & \underbrace{(s-\lambda_m)}_{\phi_m(s)} \end{pmatrix} u_{bl} = \dot{u}_{bl} - \underbrace{\begin{pmatrix} \lambda_1 & \cdots & 0 \\ \vdots & \ddots & \vdots \\ 0 & \cdots & \lambda_m \end{pmatrix}}_{\Lambda_u} u_{bl} \quad (8.25)$$

Substituting (8.5) and (8.17) into (8.25), gives

$$U_{bl} = -K_x(Ax + \tilde{B}\tilde{u}) + \Lambda_u K_x x = \underbrace{(-K_x \tilde{B})}_{(H_{\tilde{u}})_1} \tilde{u} + \underbrace{(\Lambda_u K_x - K_x A)}_{(H_x)_1} x \quad (8.26)$$

where,

$$(H_{\tilde{u}})_1 = -K_x \tilde{B} = -\begin{pmatrix} K_I & K_P \end{pmatrix} \begin{pmatrix} I_m & D_{p\,reg} \\ 0_{n \times m} & B_p \end{pmatrix} = \begin{pmatrix} \underbrace{(-K_I)}_{(H_{\tilde{u}})_{1,1}} & \underbrace{(-K_x B)}_{(H_{\tilde{u}})_{1,2}} \end{pmatrix} \neq 0 \quad (8.27)$$

represents the first $m-$rows of the control sensitivity matrix $H_{\tilde{u}}$, which in turn confirms that the relative degree of $u_{bl}$ with respect to $\tilde{u}$ is indeed one, since the system first Markov parameter from $\tilde{u}$ to $u_{bl}$ is not zero. Relations (8.26) and (8.27), yield the first $m-$rows of the sensitivity matrices $H_{\tilde{u}}$ and $H_x$, respectively.

$$(H_{\tilde{u}})_1 = (-K_I \quad -K_x B), \quad (H_x)_1 = \Lambda_u K_x - K_x A \tag{8.28}$$

Suppose that the limited output $z_{\lim}$ has a well-defined vector relative degree with respect to the system original control input $u$.

$$r_{z_{\lim}} = \left( (r_{z_{\lim}})_1 \quad \cdots \quad (r_{z_{\lim}})_m \right) \tag{8.29}$$

Then similar to (8.24),

$$Z_{\lim} = \underbrace{\begin{pmatrix} \phi_{m+1}(s) & \cdots & 0 \\ \vdots & \ddots & \vdots \\ 0 & \cdots & \phi_{2m}(s) \end{pmatrix}}_{\Phi_{z_{\lim}}(s)} z_{\lim} = \Phi_{z_{\lim}}(s) z_{\lim} = H_u u + \begin{pmatrix} 0_{m \times m} & H_{x_p} \end{pmatrix} x \tag{8.30}$$

where $\Phi_{z_{\lim}}(s) \in R^{m \times m}$ is the corresponding diagonal matrix of stable polynomials, whose orders are defined by the output relative degree vector $r_{z_{\lim}}$, component-wise. In (8.30), the control sensitivity matrix,

$$H_u = \begin{pmatrix} (C_{p\lim})_1 A_p^{(r_{z_{\lim}})_1 - 1} \\ \vdots \\ (C_{p\lim})_m A_p^{(r_{z_{\lim}})_m - 1} \end{pmatrix} B_p \in R^{m \times m} \tag{8.31}$$

is nonsingular and $H_{x_p} \in R^{m \times n_p}$ represents the original system state sensitivity matrix. In terms of the extended system dynamics (8.17), the modified output (8.30) becomes,

$$Z_{\lim} = \Phi_{z_{\lim}}(s) z_{\lim} = \underbrace{(0_{m \times m} \quad H_u)}_{(H_{\tilde{u}})_2} \tilde{u} + \underbrace{(0_{m \times m} \quad H_{x_p})}_{(H_x)_2} x \tag{8.32}$$

and shows that with respect to $\tilde{u}$, the output vector relative degree remains the same. As a consequence, the total vector relative degree of $y_{\lim}$ with respect to $\tilde{u}$ is well-defined, since the sensitivity matrix $H_{\tilde{u}}$ (8.24) is upper-triangular and nonsingular. The latter follows from the fact that the integral gain matrix $K_I$ and the control sensitivity matrix $H_u$ are both nonsingular. In this case, the CBF control augmentation component $w$ is decoupled from the AW signal $v$. Also, the inverse of $H_{\tilde{u}}$ is well-defined, upper-triangular and can be computed explicitly.

$$H_{\tilde{u}} = \begin{pmatrix} -K_I & -K_x B \\ 0_{m \times m} & H_u \end{pmatrix}, \quad H_{\tilde{u}}^{-1} = \begin{pmatrix} -K_I^{-1} & -K_I^{-1} K_x B H_u^{-1} \\ 0_{m \times m} & H_u^{-1} \end{pmatrix} \tag{8.33}$$

Furthermore, combining (8.28) and (8.32), gives the state sensitivity matrix $H_x$ in (8.22),

$$H_x = \begin{pmatrix} \Lambda_u K_x - K_x A \\ (0_{m \times m} \quad H_{x_p}) \end{pmatrix} \tag{8.34}$$

which also has a low-triangular form.

C. *Closed-Loop System Stability and Uniform Boundedness*

Similar to (4.2) and (4.3), the total extended control can be written as,

$$\tilde{u} = H_{\tilde{u}}^{-1}\left(Y_{\lim} - H_x x\right) \tag{8.35}$$

and after substituting (8.35) into (8.17), the extended closed-loop system dynamics,

$$\dot{x} = \underbrace{\left(A - \tilde{B} H_{\tilde{u}}^{-1} H_x\right)}_{\tilde{A}_{cl}} x + \tilde{B} H_u^{-1} Y_{\lim} = \tilde{A}_{cl} x + \tilde{B} H_u^{-1} Y_{\lim} \tag{8.36}$$

are guaranteed to generate uniformly bounded trajectories forward in time, as long as the corresponding closed-loop system matrix

$$\tilde{A}_{cl} = A - \tilde{B}\underbrace{\left(H_{\tilde{u}}^{-1} H_x\right)}_{K_{CBF}} = A - \tilde{B} K_{CBF} \tag{8.37}$$

is Hurwitz, where $K_{CBF}$ denotes the $((2m) \times n)-$ dimensional CBF augmentation feedback gain matrix.

### D. CBF Augmentation Feedback Gain

Based on the CBF representations (8.35), (8.37) and using the sensitivity matrices from (8.33) and (8.34), consider the corresponding CBF augmentation gain.

$$K_{CBF} = \begin{pmatrix} (K_{CBF})_v \\ (K_{CBF})_w \end{pmatrix} = H_{\tilde{u}}^{-1} H_x = \begin{pmatrix} -K_I^{-1} & -K_I^{-1} K_x B H_u^{-1} \\ 0_{m \times m} & H_u^{-1} \end{pmatrix} \begin{pmatrix} \Lambda_u K_x - K_x A \\ \begin{pmatrix} 0_{m \times m} & H_{x_p} \end{pmatrix} \end{pmatrix}$$

$$= \begin{pmatrix} -K_I^{-1}\left(\Lambda_u K_x - K_x A\right) - K_I^{-1} K_x \begin{pmatrix} 0_{m \times m} & B H_u^{-1} H_{x_p} \end{pmatrix} \\ \begin{pmatrix} 0_{m \times m} & H_u^{-1} H_{x_p} \end{pmatrix} \end{pmatrix} \tag{8.38}$$

$$= \begin{pmatrix} -K_I^{-1}\Lambda_u K_x + K_I^{-1} K_x \begin{pmatrix} 0_{m \times m} & A - B H_u^{-1} H_{x_p} \end{pmatrix} \\ \begin{pmatrix} 0_{m \times m} & H_u^{-1} H_{x_p} \end{pmatrix} \end{pmatrix}$$

Since,

$$K_I^{-1} \Lambda_u K_x = \begin{pmatrix} K_I^{-1} \Lambda_u K_I & K_I^{-1} \Lambda_u K_P \end{pmatrix} \tag{8.39}$$

then

$$K_{CBF} = \begin{pmatrix} (K_{CBF})_v \\ (K_{CBF})_w \end{pmatrix} = \begin{pmatrix} -\begin{pmatrix} K_I^{-1}\Lambda_u K_I & K_I^{-1}\Lambda_u K_P \end{pmatrix} + K_I^{-1} K_x \begin{pmatrix} 0_{m \times m} & A - B H_u^{-1} H_{x_p} \end{pmatrix} \\ \begin{pmatrix} 0_{m \times m} & H_u^{-1} H_{x_p} \end{pmatrix} \end{pmatrix}$$

$$= \begin{pmatrix} -K_I^{-1}\Lambda_u K_I & K_I^{-1}\left(\Lambda_u K_P + K_x\left(A - B H_u^{-1} H_{x_p}\right)\right) \\ 0_{m \times m} & H_u^{-1} H_{x_p} \end{pmatrix} \tag{8.40}$$

gives an explicit representation of the CBF augmentation controller gains.

$$K_{CBF} = \begin{pmatrix} (K_{CBF})_v \\ (K_{CBF})_w \end{pmatrix} = \begin{pmatrix} -K_I^{-1}\Lambda_u K_I & K_I^{-1}\left(\Lambda_u K_P + K_x\left(A - B H_u^{-1} H_{x_p}\right)\right) \\ 0_{m \times m} & H_u^{-1} H_{x_p} \end{pmatrix} \tag{8.41}$$

Clearly, (8.41) implies that the CBF AW matrix of gains $(K_{CBF})_v$ adds a negative-definite damping term $(K_I^{-1}\Lambda_u K_I)$ into the integrator dynamics, while the second matrix row of gains $(K_{CBF})_w$ in (8.41) due to the CBF augmentation signal $w$, shows that the baseline controller gains are replaced by the output feedback DI-like term $\left[-(H_u^{-1}H_{x_p})x_p\right]$. That is the "cancellation" nature of the CBF method.

E.  *System Control Block-Diagram*

Figure 6 shows the system block-diagram with a baseline state feedback PI servo-controller and the min-norm optimal piece-wise linear state feedback CBF augmentation for enforcing the desired operational min/max constraints that are imposed on the system total input and on the selected output.

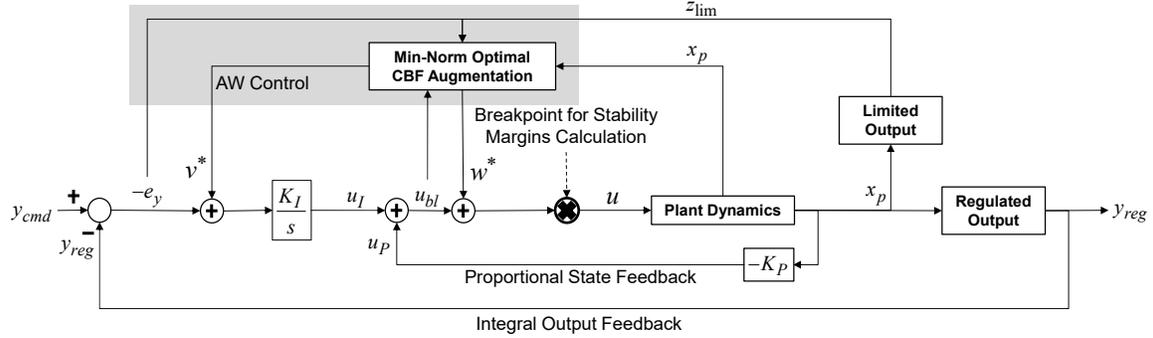

**Figure 6 Closed-loop system block-diagram with min-norm optimal CBF servo-control augmentation**

By design, the CBF servo-control augmentation logic enforces soft constraints on the total control input, that is no hard constraints, that would be represented by the sat-function, are required. Using hard saturation logic is standard in control applications and it can be incorporated into the controller block-diagram for practical purposes to mitigate potentially undesirable effects due to numerical implementation of the algorithm. The selected limited output is also subject to soft constraints and requires no explicit hard saturation. In addition to soft-constrained total control and limited output signals, the CBF augmentation solution adds an anti-windup protection with respect to the controller integrator state components, keeping them uniformly bounded during input/output soft saturation events.

F.  *Relative Stability and Margins Evaluation*

For relative stability analysis and margin calculations, the extended total control solution (8.23) can be rewritten in the form similar to (5.5),

$$\tilde{u} = \begin{pmatrix} v^* - y_{cmd} \\ u_{bl} + w^* \end{pmatrix} = \underbrace{\begin{pmatrix} -y_{cmd} \\ u_{bl} \end{pmatrix}}_{\tilde{u}_{bl}} - H_{\tilde{u}}^{-1}\tilde{\delta}(x)\left(H_x x + H_{\tilde{u}}\tilde{u}_{bl} - \alpha_\pi y_{\lim}^{\min/\max}\right)$$

$$= \underbrace{\left(I_{(2m)} - H_{\tilde{u}}^{-1}\tilde{\delta}(x)H_{\tilde{u}}\right)\tilde{u}_{bl}}_{\text{Scaled Extended Baseline Control}} - \underbrace{H_{\tilde{u}}^{-1}\tilde{\delta}(x)H_x x}_{\text{CBF Feedback}} + \underbrace{H_{\tilde{u}}^{-1}\tilde{\delta}(x)\alpha_\pi y_{\lim}^{\min/\max}}_{\text{CBF Command}}$$

(8.42)

with the $((2m)\times(2m))$– dimensional non-negative binary-valued diagonal matrix,

$$\tilde{\delta}(x) = \begin{pmatrix} \tilde{\delta}_1(x) & \cdots & 0 \\ \vdots & \ddots & \vdots \\ 0 & \cdots & \tilde{\delta}_{(2m)}(x) \end{pmatrix} \in R^{(2m)\times(2m)}$$

$$\tilde{\delta}_i(x) = \begin{cases} 1, & \text{if }\left[\left(-H_x x - H_{\tilde{u}}\tilde{u}_{bl} + \alpha_\pi y_{\lim}^{\min}\right)_i > 0\right] \vee \left[\left(H_x x + H_{\tilde{u}}\tilde{u}_{bl} - \alpha_\pi y_{\lim}^{\max}\right)_i > 0\right] \\ 0, & \text{otherwise} \end{cases}$$

(8.43)

and using the corresponding component-wise output command in the form of (5.6).

$$\left(y_{\lim}^{\min/\max}\right)_i = \begin{cases} \left(y_{\lim}^{\min}\right)_i, & \text{if } \left(-H_x x - H_{\tilde{u}} \tilde{u}_{bl} + \alpha_\pi y_{\lim}^{\min}\right)_i > 0 \\ \left(y_{\lim}^{\max}\right)_i, & \text{if } \left(H_x x + H_{\tilde{u}} \tilde{u}_{bl} - \alpha_\pi y_{\lim}^{\max}\right)_i > 0, \quad i = 1,\ldots,(2m) \\ 0, & \text{otherwise} \end{cases} \quad (8.44)$$

Within the servo-control design framework, gain and phase margins need to be analyzed at the total control input breakpoint (Figure 6), where the baseline control $u_{bl}$ and the CBF augmentation signal $w^*$ are added to form the total control input $u$ into the system. For this analysis, all active AW augmentation loops due to $v^*$ should be closed, since the integrator dynamics represent a known part of the total controller.

The system loop gain transfer function matrix $L_u(s)$ can be computed similarly to (5.8), with the total servo-controller in the form of (8.42), while zeroing out command terms $y_{cmd}$ and $y_{\lim}^{\min/\max}$.

$$u_{out} = \begin{pmatrix} 0_{m\times m} & I_m \end{pmatrix}\left(I_{(2m)} - H_{\tilde{u}}^{-1}\tilde{\delta}H_{\tilde{u}}\right)\underbrace{\tilde{u}_{bl}}_{\begin{pmatrix} 0_{m\times n} \\ -K_x x \end{pmatrix}} - \begin{pmatrix} 0_{m\times m} & I_m \end{pmatrix}H_{\tilde{u}}^{-1}\tilde{\delta}H_x x$$

$$= -\left(\left(I_m - \underbrace{\begin{pmatrix} 0_{m\times m} & I_m \end{pmatrix}H_{\tilde{u}}^{-1}\tilde{\delta}}_{\delta}H_{\tilde{u}}\right)\begin{pmatrix} 0_{m\times m} \\ I_m \end{pmatrix}K_x x + \delta H_x x\right) \quad (8.45)$$

$$= -\underbrace{\left[\left(I_m - \delta H_{\tilde{u}}\begin{pmatrix} 0_{m\times m} \\ I_m \end{pmatrix}\right)K_x + \delta H_x\right](sI_n - A)^{-1}B}_{\text{Loop Gain}: L_u(s;\delta)}u_{in} = -L_u(s;\delta)u_{in}$$

In this case, SISO and MIMO margins at the system input breakpoint are defined based on the resulting $(m\times m)$– dimensional loop gain transfer function matrix, parameterized with the binary-valued matrix $\delta$,

$$L_u(s;\delta) = \left(\left(I_m - \delta H_{\tilde{u}}\begin{pmatrix} 0_{m\times m} \\ I_m \end{pmatrix}\right)K_x + \delta H_x\right)(sI_n - A)^{-1}B \quad (8.46)$$

where

$$\delta = \begin{pmatrix} 0_{m\times m} & I_m \end{pmatrix}H_{\tilde{u}}^{-1}\tilde{\delta} = \begin{pmatrix} 0_{m\times m} & I_m \end{pmatrix}\begin{pmatrix} -K_I^{-1} & -K_I^{-1}K_x B H_u^{-1} \\ 0_{m\times m} & H_u^{-1} \end{pmatrix}\tilde{\delta} = \begin{pmatrix} 0_{m\times m} & H_u^{-1} \end{pmatrix}\tilde{\delta} \quad (8.47)$$

represents the matrix of active CBF constraints on the system output.

## 9. Flight Control Design and Simulation Trade Study

Consider the roll-yaw dynamics representative of a mid-size aircraft, ([3], Section 14.8, pp. 622–626).

$$\underbrace{\begin{pmatrix} \dot{\beta} \\ \dot{p}_s \\ \dot{r}_s \end{pmatrix}}_{\dot{x}_p} = \underbrace{\begin{pmatrix} \frac{Y_\beta}{V_0} & \frac{Y_{p_s}}{V_0} & \frac{Y_{r_s}}{V_0}-1 \\ L_\beta & L_{p_s} & L_{r_s} \\ N_\beta & N_{p_s} & N_{r_s} \end{pmatrix}}_{A_p}\underbrace{\begin{pmatrix} \beta \\ p_s \\ r_s \end{pmatrix}}_{x_p} + \underbrace{\begin{pmatrix} \frac{Y_{\delta_{ail}}}{V_0} & \frac{Y_{\delta_{rud}}}{V_0} \\ L_{\delta_{ail}} & L_{\delta_{rud}} \\ N_{\delta_{ail}} & N_{\delta_{rud}} \end{pmatrix}}_{B_p}\underbrace{\begin{pmatrix} \delta_{ail} \\ \delta_{rud} \end{pmatrix}}_{u}$$

The system state $x_p$ includes the aircraft sideslip angle $\beta$ (rad), as well as the vehicle stability axis roll and yaw rates (rad/sec), $p_s$ and $r_s$. The control input $u$ is represented by the aileron and the rudder deflections (rad), $\delta_a$ and $\delta_r$. The regulated output of interest consists of the aircraft roll rate $p_s$ (rad/sec) and the lateral load factor $N_y$ (g-s), where $g = 32.174$ is the gravitational acceleration (ft/sec²).

$$y_{reg} = \begin{pmatrix} p_s & N_y \end{pmatrix}^T = \underbrace{\begin{pmatrix} 0 & 1 & 0 \\ \frac{Y_\beta}{g} & \frac{Y_{p_s}}{g} & \frac{Y_{r_s}}{g} \end{pmatrix}}_{C_{preg}} x_p + \underbrace{\begin{pmatrix} 0 & 0 \\ \frac{Y_{\delta_{ail}}}{g} & \frac{Y_{\delta_{rud}}}{g} \end{pmatrix}}_{D_{preg}} u = C_{preg} x_p + D_{preg} u$$

The aircraft model data are computed using numerical linearization with respect to a 1g-level flight trim (i.e., equilibrium) at the selected flight conditions.

$$V_0 = 717.17 \left(\frac{ft}{sec}\right), \quad Alt = 25000 (ft), \quad \alpha = 4.5627 (\deg)$$

$$A_p = \begin{pmatrix} -0.11794 & 0.00085 & -1.0001 \\ -7.0113 & -1.4492 & 0.22059 \\ 6.3035 & 0.06511 & -0.41172 \end{pmatrix}, \quad B_p = \begin{pmatrix} 0 & 0.015257 \\ -7.9662 & 2.6875 \\ 0.60926 & -2.3577 \end{pmatrix}$$

$$C_{preg} = \begin{pmatrix} 0 & 1 & 0 \\ -2.6049 & 0.018724 & 0.067695 \end{pmatrix}, \quad D_{preg} = \begin{pmatrix} 0 & 0 \\ 0 & 0.33698 \end{pmatrix}$$

A baseline LQR PI controller is designed without operational limits, using the integrated output tracking error dynamics,

$$\dot{e}_I = y_{reg} - y_{cmd} = \begin{pmatrix} p_s - p_{s\,cmd} \\ N_y - N_{y\,cmd} \end{pmatrix}$$

and the following LQR weights.

$$Q_{lqr} = \text{diag}(1.025 \quad 1.0289 \quad 0 \quad 0 \quad 1.6021), \quad R_{lqr} = \text{diag}(1 \quad 0.49129)$$

Figure 7 shows adequate closed-loop system tracking performance due to external step-input commands.

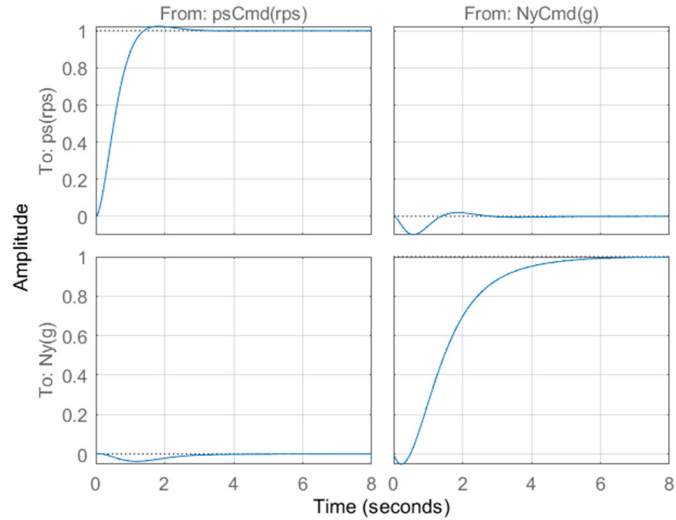

Figure 7 Closed-loop system tracking performance with unconstrained baseline LQR PI controller

Because of the tracking error integrators, dynamics of the two regulated outputs are almost decoupled. Figure 8 shows the LQR PI loop gains at the system input break-points, computed one at a time with and without an actuator model ("subsystems").

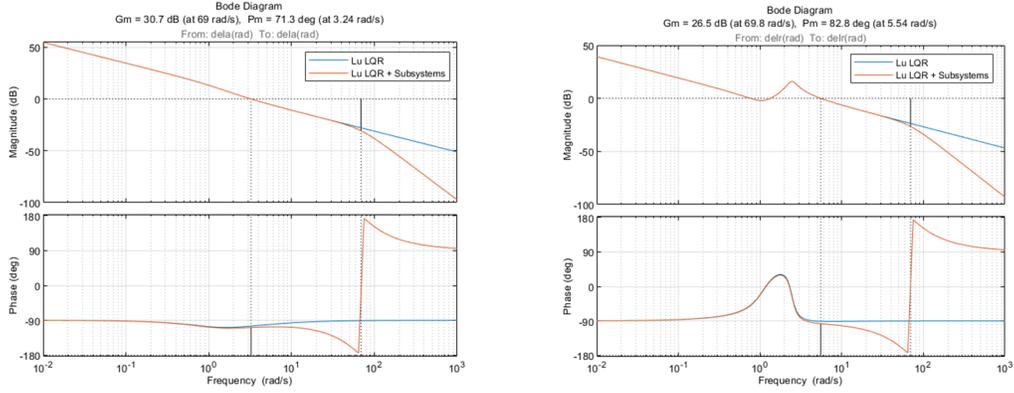

**Figure 8  Loop gains with unconstrained baseline LQR PI controller**

These data confirm satisfactory robustness and command tracking characteristics of the baseline controller, when it is operating without min/max limits.

For this case study, the selected limited output includes the aircraft roll rate and the sideslip angle.

$$z_{\lim} = \begin{pmatrix} p_s \\ \beta \end{pmatrix} = \underbrace{\begin{pmatrix} 0 & 1 & 0 \\ 1 & 0 & 0 \end{pmatrix}}_{C_p} x_p + \underbrace{\begin{pmatrix} 0 & 0 \\ 0 & 0 \end{pmatrix}}_{D_p} u$$

The limited output vector relative degree is $r = (1, 2)$ with respect to the system control input. It can be verified that the corresponding input-output matrix $H_w$ is nonsingular.

Consider the closed-loop system response using the unconstrained baseline LQR PI controller, which is tested with a series of $(\pm 4\,\deg/\sec)$ step-input commands in $p_{s\,cmd}$ and $(\pm 0.03\,g)$ command in $N_{y\,cmd}$. In practical applications, such a test would be representative of uncoordinated turn capabilities, as shown in Figure 9.

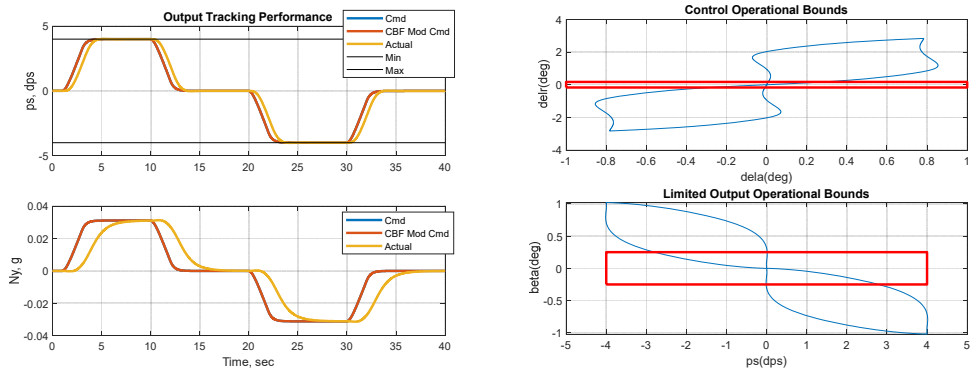

**Figure 9  Simulation with unconstrained baseline LQR PI controller**

A min-norm optimal augmentation controller is constructed using the equations from Table 1. The CBF design starts with the selection of a diagonal positive-definite $(4 \times 4)$–matrix,

$$\alpha_\pi = \mathrm{diag}\begin{pmatrix} 80 & 8 & 40 & 40 \end{pmatrix} > 0$$

whose diagonal elements (8.22),

$$c_{i0} = \prod_{j=1}^{r_i}(-\lambda_{ij}) > 0, \quad \forall i = 1,\ldots,(2m=4)$$

are products of the selected real negative eigenvalues of the stable polynomials (2.6),

$$\phi_i(s) = \prod_{j=1}^{r_i}(s - \lambda_{ij}) = \sum_{j=0}^{r_i} c_{ij} s^j, \quad \forall i = 1,\ldots,(2m=4)$$

that define $Y_{\lim}$. In this case, they are: aileron, rudder, roll rate and sideslip signals, correspondingly. Orders $r_i$ of the polynomials are equal to the individual output relative degrees: $(1\ 1\ 1\ 2)$. One can select diagonal elements of $\alpha_\pi$ first and then compute the corresponding eigenvalues. Note that large positive values for $c_{i0}$ decrease CBF conservatism near min/max boundaries.

For simulation and testing purposes, the aileron and the rudder position limits are set to $(\pm 1\,\text{deg})$ and $\left(\pm\frac{1}{6}\,\text{deg}\right)$, correspondingly. In addition, the roll rate and the sideslip limits are $(\pm 4\,\text{deg})$ and $(\pm 0.25\,\text{deg})$. Selection of these small operational limits allows to demonstrate efficiency of the CBF control augmentation.

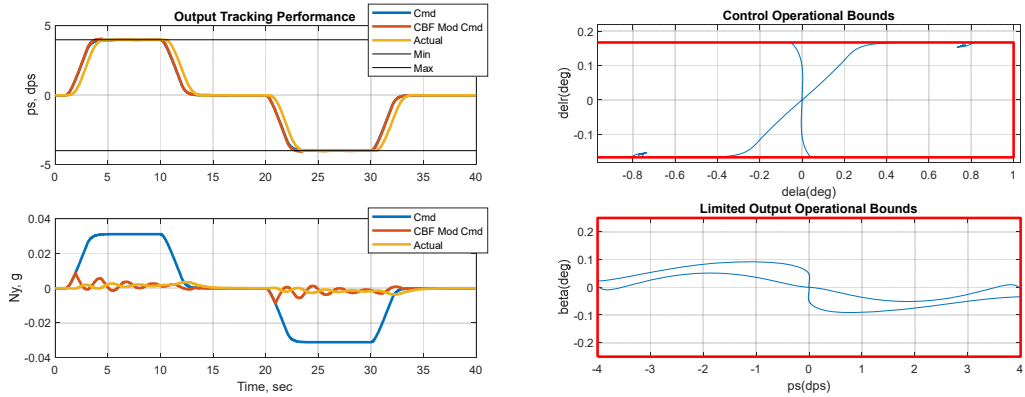

**Figure 10  Simulation with constrained LQR PI controller and min-norm CBF-based control augmentation, in the presence of operational limits**

As seen from the data, the roll rate tracks its commanded value (left upper plot), while the aileron and the rudder channels saturate most of the time, (right upper phase plot). The saturation of the control surfaces, drives the achievable side acceleration to become much smaller than the command (left bottom plot). Control position and rate data for the same test are shown in Figure 11.

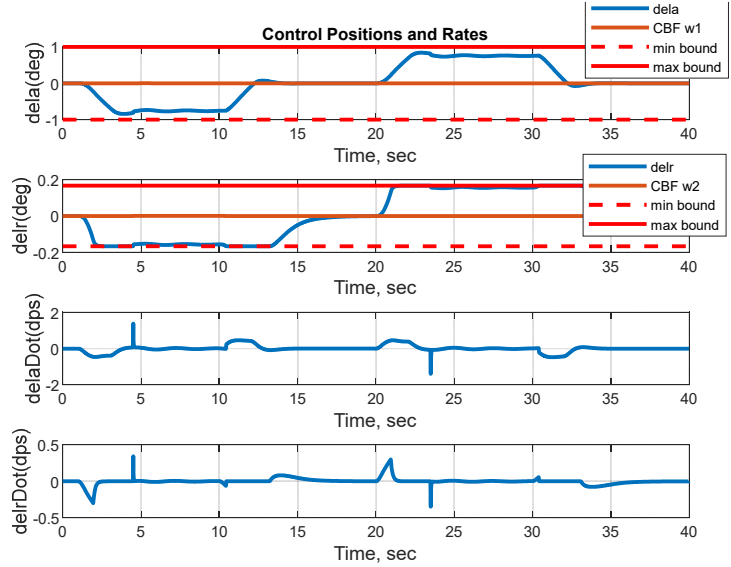

**Figure 11 Control positions and rates during uncoordinated turn with constrained LQR PI controller and min-norm CBF-based control augmentation**

The observed control activity is within reasonable actuation bounds. The system states, including controller tracking error integrators, are shown in Figure 12.

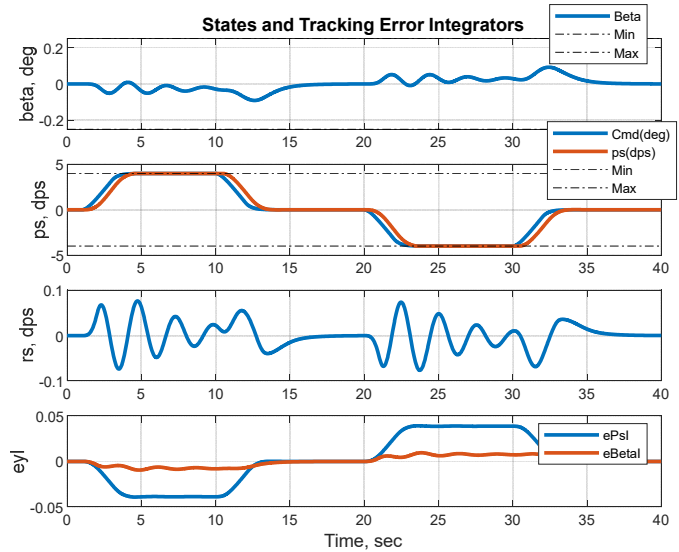

**Figure 12 System states during uncoordinated turn with constrained LQR PI controller and min-norm CBF-based control augmentation**

The important feature of the CBF augmentation design is the anti-windup protection for the roll rate and sideslip tracking error integrators $\left(e_{p_s I}, e_{\beta I}\right)$ during saturation events. All signals have acceptable transients.

Figure 13 shows MIMO gain and phase margins versus CBF configuration number, as defined by the loop gain transfer function (5.11), computed at the system input breakpoint, with all of the CBF anti-windup augmentation loops closed.

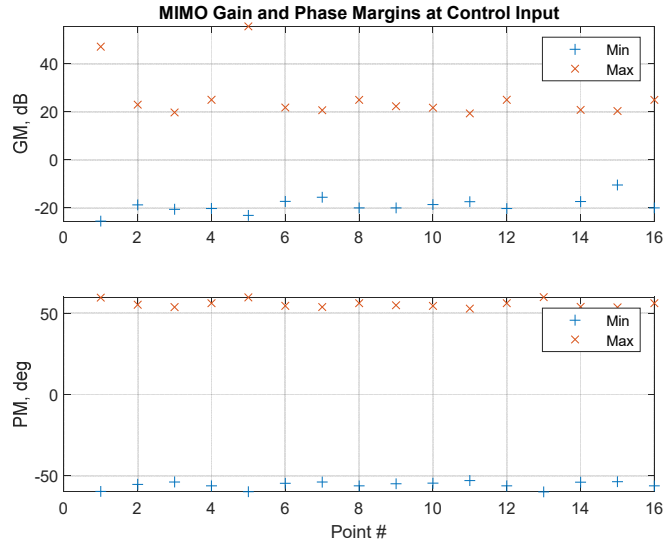

**Figure 13** MIMO Gain and phase margins for (LQR PI + CBF) controller at the system input breakpoint, with CBF AW loops closed

As seen from the figure, all possible combinations of active operational constraints show that the system has large and definitely acceptable stability margins in the MIMO sense.

Overall, simulation test and analysis data show potential benefits of the developed control augmentation solution for flight critical control applications, such as aircraft primary flight control systems. Specifically, this technology can be used to design output and control limiters to enforce operational limits for aerial vehicles.

## 10. Conclusions

In this paper, a formal control augmentation design method is developed for MIMO LTI systems with a baseline PI servo-controller subject to box constraints that represent the desired operational limits imposed on the system control input and on a selected output. The design is based on the Nagumo Theorem [6], the Comparison Lemma, and the min-norm optimal controllers [4] with QP optimization [5]. The design connections to CBF-based methods [8], [9] are discussed. The developed solution also provides an anti-windup protection for the controller integrator state and it enforces component-wise soft min/max constraints on the total control command, as well as on the selected output.